\documentclass[12pt,amssymb]{article} 

\usepackage{mathrsfs}
\usepackage{amssymb}
\usepackage[dvips]{graphicx}

\newcommand{\newsection}[1]{
\addtocounter{section}{1} \setcounter{equation}{0}
\setcounter{subsection}{0} \addcontentsline{toc}{section}{\protect
\numberline{\arabic{section}}{{\rm #1}}} \vglue .6cm \pagebreak[3]
\noindent{ \bf  \thesection. #1}\nopagebreak[4]\par\vskip .3cm}
\newcommand{\newsubsection}[1]{
\addtocounter{subsection}{1}\setcounter{subsubsection}{0}
\addcontentsline{toc}{subsection}{\protect
\numberline{\arabic{section}.\arabic{subsection}}{#1}} \vglue .4cm
\pagebreak[3] \noindent{\it \thesubsection.
#1}\nopagebreak[4]\par\vskip .3cm}

\makeatletter
\newcommand{\seclabel}[1]{%
  \@bsphack
  \protected@write\@auxout{}%
     {\string\newlabel{#1}{{\thesection}{\thepage}}}
  \@esphack
  }
\newcommand{\subseclabel}[1]{%
  \@bsphack
  \protected@write\@auxout{}%
     {\string\newlabel{#1}{{\thesubsection}{\thepage}}}
  \@esphack
  }
\newcommand{\tablabel}[1]{%
  \@bsphack
  \protected@write\@auxout{}%
     {\string\newlabel{#1}{{\arabic{tabnum}}{\thepage}}}
  \@esphack
  }
\makeatother

\renewcommand{\theequation}{\thesection.\arabic{equation}}

\newlength{\extraspace}
\newlength{\extraspaces}
\newcounter{dummy}
\newcommand{\bc}{\begin{center}}
\newcommand{\ec}{\end{center}}
\newcommand{\be}{\begin{equation}
\addtolength{\abovedisplayskip}{\extraspaces}
\addtolength{\belowdisplayskip}{\extraspaces}
\addtolength{\abovedisplayshortskip}{\extraspace}
\addtolength{\belowdisplayshortskip}{\extraspace}}
\newcommand{\ee}{\end{equation}}

\newcommand{\ba}{\begin{eqnarray}
\addtolength{\abovedisplayskip}{\extraspaces}
\addtolength{\belowdisplayskip}{\extraspaces}
\addtolength{\abovedisplayshortskip}{\extraspace}
\addtolength{\belowdisplayshortskip}{\extraspace}}
\newcommand{\ea}{\end{eqnarray}}

\newcommand{\is}{& \!\! = \!\! &}
\newcommand{\ban}{\begin{eqnarray*}
\addtolength{\abovedisplayskip}{\extraspaces}
\addtolength{\belowdisplayskip}{\extraspaces}
\addtolength{\abovedisplayshortskip}{\extraspace}
\addtolength{\belowdisplayshortskip}{\extraspace}}
\newcommand{\ean}{\end{eqnarray*}}
\newcommand{\baa}{
\addtocounter{equation}{1} \setcounter{dummy}{\value{equation}}
\setcounter{equation}{0}
\renewcommand{\theequation}{\thesection.\arabic{dummy}\alph{equation}}
\begin{eqnarray}
\addtolength{\abovedisplayskip}{\extraspaces}
\addtolength{\belowdisplayskip}{\extraspaces}
\addtolength{\abovedisplayshortskip}{\extraspace}
\addtolength{\belowdisplayshortskip}{\extraspace}}
\newcommand{\eaa}{
\end{eqnarray}
\setcounter{equation}{\value{dummy}}
\renewcommand{\theequation}{\thesection.\arabic{equation}}}

\newcommand{\fraam}[1]{\fbox{$\strut \qquad #1 \qquad$}}
\input epsf

\newcounter{fignum}
\newcounter{tabel}

\newcounter{tabnum}
\newcommand{\vev}[1]{\left\langle #1\right\rangle}

\newcommand{\half}{\frac{1}{2}}
\newcommand{\del}{\partial}
\newcommand{\delb}{\bar{\del}}
\newcommand{\eol}{\nonumber \\}

\newcommand{\cO}{{\cal O}}
\newcommand{\Ext}{{\rm Ext}}
\newcommand{\Hom}{{\rm Hom}}

\begin{document}

%
%

\begin{flushright}
October 2023\\
\end{flushright}
\vspace{2cm}

\thispagestyle{empty}

\begin{center}
{\Large\bf The $M$-Theory Three-Form and Singular Geometries
 \\[13mm] }

{\sc Ron Donagi \& Martijn Wijnholt}\\[2.5mm]
{\it Department of Mathematics, University of Pennsylvania \\
Philadelphia, PA 19104-6395, USA}\\[15mm]

Abstract:
\end{center}
While $M$- and $F$-theory compactifications describe a much larger class of vacua than perturbative string compactifications, 
they typically need singularities to generate non-abelian gauge fields and charged matter.
The physical explanation involves $M2$-branes wrapped on
vanishing cycles. Here we seek an alternative explanation that could address outstanding issues such as the description of 
nilpotent branches, stability walls, frozen singularities and so forth. 
To this end we use a model in which the three-form is related to the Chern-Simons form of a bundle.
The model has a one-form non-abelian gauge symmetry which normally eliminates all the degrees of freedom associated to the bundle.
However by restricting the transformations to preserve the bundle along the vanishing cycles, we may get new degrees of freedom associated to singularities,
without appealing to wrapped $M2$-branes. The analysis can be simplified by gauge-fixing the one-form symmetry using higher-dimensional instanton equations.
We explain how this mechanism leads to the natural emergence of phenomena such as enhanced ADE gauge symmetries, nilpotent branches, charged matter fields and their holomorphic couplings.

\newpage

\renewcommand{\Large}{\normalsize}

\tableofcontents

\newpage

\newsection{Introduction}

The goal of this paper is to improve our understanding of  $F$-theory (or $M$-theory) compactifications on singular spaces. There are usually some extra light degrees of freedom associated to the singularities, and our objective will be to fit these together with the usual 
supergravity degrees of freedom in the framework of deformation theory. This has some perhaps surprising connections with a range of topics.
It leads to interesting questions about a proper description of the $M$-theory three-form tensor field in the presence of defects. It can also be related,
at least in the context we study, to the derived geometry of sheaves and bundles on a Calabi-Yau four-fold. But let us start with some motivation for such an approach.

One of the dominant themes in the study of string compactifications is the relation between
mathematical structures in the compactification data and the low energy effective Lagrangian
in the uncompactified dimensions. A perturbative string compactification is characterized by some
algebraic or differential geometric structure, and the low energy effective action encapsulates the deformation
theory of this structure. Typically symmetries of the compactification data get related to gauge fields,
deformations
get related to matter fields and moduli, and obstructions
to deformations get mapped to interaction terms in the low energy effective theory.

Consider as a proto-typical example the heterotic string compactified on a Calabi-Yau three-fold $Z$ with a holomorphic
$E_8\times E_8$ bundle. Focussing on one of the $E_8$'s at a time, the corresponding bundle $V_{E_8}$ breaks the ten-dimensional $E_8$ 
gauge symmetry to a smaller group,
and generates charged matter fields in the process. Symmetries of the $E_8$ bundle describe the unbroken part of the gauge group, 
and lead to four-dimensional gauge fields.
Deformations of $TZ$, $T^*Z$ and the $E_8$ bundle lead to moduli and matter fields. Obstructions are encoded
in the holomorphic Chern-Simons superpotential:
\be\label{hCS}
W(A) \ = \ \int_Z \Omega^{3,0} \wedge \omega_{CS}(A)
\ee
where $A$ is a connection on $V_{E_8}$.
By differentiating three times with respect to bundle deformations, we find that the Yukawa couplings of four-dimensional matter fields are computed by the triple products:
\be\label{CS_Yukawa}
H^1(Z, V_{E_8}) \times H^1(Z, V_{E_8})\times H^1(Z, V_{E_8})\ \to\ {\bf C}
\ee
In fact the holomorphic Chern-Simons superpotential encodes a whole $A_\infty$ algebra,
with higher order Massey products describing higher order interactions in the low energy effective theory \cite{Witten:1992fb,Merkulov_homotopy,Polishchuk_HM}.
Similar structures may be found in type II compactifications with $A$- or $B$-branes. There is by now a large literature, but see eg. \cite{Kontsevich:2000yf,Lazaroiu:2001nm,Kajiura:2005sn,Aspinwall:2009isa} for some early work.

Although this general framework looks pretty reasonable, it seems to be insufficient to describe situations where the compactification manifold has singularities, and there are fields in the low energy effective theory that originate from non-perturbative solitons wrapping vanishing cycles of the singularity. The degrees of freedom from these solitons are essentially added to the supergravity fields \cite{Strominger:1995cz}, and even though they are in some sense classical (since they are not suppressed by any coupling constant when the vanishing cycles have collapsed to zero size), in our current understanding they don't come from deformation of some geometric structure of the compactification. Prominent examples are compactifications of $M$- or $F$-theory, where such singularities are essential to generate non-abelian gauge theories.

For concreteness let us focus on one of the most popular models, namely $F$-theory compactified on an eight-dimensional manifold $Y$ with ${\sf G}_4$-flux
\cite{Vafa:1996xn,Morrison:1996na,Morrison:1996pp,Becker:1996gj}.
In the $F$-theory picture, non-abelian gauge fields and charged matter instead arise from quantized $M2$-branes wrapped on exceptional ${\bf P}^1$'s. Effective field theory and duality with the heterotic string suggest there is again
an underlying deformation theoretic picture, governed by a superpotential. But from the $F$-theory perspective it is rather unclear what is being deformed, since the degrees of freedom from wrapped $M2$-branes are not represented by geometric data of the compactification. This suggests looking for an alternative approach that avoids the mismatch.

This brings us to the main point of the present paper. We will argue that not only can we recover the complete deformation theoretic picture expected from dualities, we can actually do so just from supergravity, that is we will have no need to invoke $M2$-branes wrapped on vanishing cycles. However the key is to pay special attention to 
the three-form field when singularities are present.

To this end we consider a model introduced in
\cite{Diaconescu:2003bm}, henceforth referred to as the DFM model, in which the three-form is related to the Chern-Simons form of a bundle (following an earlier suggestion in \cite{Witten:1996md}). 
This leads us to add a holomorphic non-abelian bundle on the four-fold, whose Chern-Simons form reproduces the given three-form. The model has a one-form gauge symmetry $A \to A + \psi$, corresponding to arbitrary shifts of the connection on the bundle. Thus on a smooth manifold this model will just give us back the usual degrees of freedom associated to the three-form field, and so doesn't appear to be very interesting. However in the presence of defects we have to worry about boundary conditions, which can give rise to new degrees of freedom because certain types of gauge transformations may be disallowed.
And indeed in the presence of singularities something interesting happens.
Precisely when we expect new degrees of freedom from $M2$-branes wrapped on vanishing cycles, the non-abelian bundle
detects new symmetries and deformations, which are described by new generators of the Dolbeault cohomology groups of the bundle. To our knowledge, in its simplest form these deformations were first discussed in \cite{FM} and subsequently more systematically in \cite{ChenLeung}. Furthermore these extra generators are special in that they are not smooth on the singular space, i.e. they have non-trivial behaviour over the exceptional cycles. Hence we conjecture that these modes are not valid one-form gauge symmetry transformations and actually describe physical degrees of freedom. In the context of heterotic/$F$-theory duality, this is also confirmed by restricting these modes to the log boundary, which reproduces the usual modes responsible for 
non-abelian gauge symmetry enhancement in the heterotic string.

Now DFM didn't mention anything about holomorphic structures on bundles or Dolbeault cohomology, so we will have to explain why these are relevant things to look at. The reasoning is as follows. The superpotential on a Calabi-Yau four-fold is invariant under the one-form topological gauge symmetry of the DFM model, which kills all the local non-abelian degrees of freedom. In order to simplify things, one may consider gauge fixing this one-form gauge symmetry, leaving at most a finite dimensional set of gauge transformations.
A convenient way to do this using instanton equations was carried out in
\cite{Baulieu:1997jx,Baulieu:1997em}, and after adding the gauge fixing terms one finds that the action for the gauge fields is essentially given by a holomorphic twist of the eight-dimensional
supersymmetric gauge theory on the Calabi-Yau four-fold.

The resulting gauge fixed theory localizes on connections that satisfy a complex
anti-self-duality (ASD) equation, which is closely related to the problem of putting holomorphic structures on bundles. Thus we are now back in a more-or-less standard deformation theoretic setting, even one that can frequently be tackled with algebraic geometry, namely the deformation theory of pairs $(Y,{\cal W})$ where $Y$ is a complex manifold and ${\cal W}$ is a holomorphic bundle. The deformation theory of the complex ASD equations on a fixed four-fold $Y$ has been elucidated in recent years \cite{Brav:2013,Cao:2014bca,Borisov:2015vha}, which is helpful in this regard and gives geometric meaning to the $F$-theory superpotential. However in our setting
there are some non-standard features. As already mentioned, while in conventional compactifications one is interested in all bundle deformations, here we need to mod out by deformations that preserve the bundle along the defect locus, because those are interpreted as remnant one-form gauge transformations in the DFM model that survived the gauge fixing.
Only conventional three-form deformations and deformations that change the bundle along the defect locus are considered physical.
Also, the natural stability criterion differs from slope stability, and has the property that we get stability walls only when the vanishing cycles have shrunk to zero size. This avoids a contradiction with the description by wrapped $M2$-branes, which requires finite size vanishing cycles and sees no stability walls.  
However the latter is not important for the holomorphic
questions that are the focus of the present paper.

In our approach, the rules for deriving the low energy effective action of an $M$-theory or $F$-theory compactification necessarily differ from the existing ones. For example, a classical result in the conventional approach is that the number of $U(1)$ gauge symmetries in the low energy effective theory is counted by degree two cohomology classes of the (resolved) Calabi-Yau. But in our approach the results may differ because the bundle need not respect these gauge symmetries. Indeed there may be physically distinct vacua associated to the same singular geometry, and such refinements of the rules involving the bundle are necessary in order to be able to distinguish between them. 

While we believe the present work gives new insight into the phenomenon of non-abelian gauge symmetry enhancement in $M/F$-theory, there are are clearly still a number of loose ends that one would like to understand better, some of which we point out in the text. For another relevant puzzle, see \cite{Freed:2019sco}. Nevertheless we hope to demonstrate that it is a useful point of view, by touching on a few applications like nilpotent branches, stability walls, Yukawa couplings and the relation between the $F$-theory and heterotic superpotentials. These items were previously hard to understand or simply out of reach with the wrapped $M2$-brane description.

This work is part of a larger project. The first part appeared in \cite{ADE_Transform} and was concerned with a construction of certain tautological bundles using a Fourier-Mukai type transform.
A brief review of this paper is included in section \ref{ADE_sheaves_bundles}.
The present paper focuses on the relation between bundles and the three-form, the (holomorphic) TQFT governing sheaves or bundles on a Calabi-Yau four-fold, and explains the physical relevance of tautological bundles promised in \cite{ADE_Transform}. Further material may appear elsewhere.

Previous attempts at describing wrapped $M2$-brane degrees of freedom with geometric data on an $F$-theory compactification
have appeared in several papers. To our knowledge, the first version of this idea appeared in 
\cite{Donagi:2011jy}, where  we argued using heterotic/$F$-theory duality that the Deligne cohomology associated to a singular $F$-theory
compactification already knows about the degrees of freedom usually ascribed to wrapped $M2$-branes. 
In \cite{Marsano:2012bf} the authors suggested to incorporate some extra data in $F$-theory in terms of a line bundle supported on a divisor in $Y$.
This looks similar to the sheaf ${\cal F}$ appearing in (\ref{four_boxes}).
In \cite{Collinucci:2014taa}
the proposal was to add a matrix factorization (which can be interpreted as a type of sheaf) to the singular
version of $Y$ to describe wrapped membrane states. Again there seem to be some similarities with the sheaf ${\cal F}$. In \cite{Anderson:2013rka} the idea of looking a the Deligne cohomology associated to a singular manifold  was expanded on by bringing in limiting mixed Hodge structures.

\newpage

\newsection{ADE bundles, ADE sheaves and the three-form}
\seclabel{ADE_sheaves_bundles}

In this section we give a overview of ADE bundles and their relation to the three-form field in
$M$-theory. We include some material from \cite{ADE_Transform,MF_Extension,F_Fluxes,F_Stability}.

\newsubsection{Non-abelian model for the three-form}
\subseclabel{DFM_model}

Motivated by earlier work of \cite{Witten:1996md}, the authors of \cite{Diaconescu:2003bm} proposed a model for the three-form in terms of non-abelian differential cohomology. We will
frequently refer to this as the DFM model for the three-form. Morally speaking, the idea is to capture the topological data of the three-form by the Chern-Simons form of a gauge field,
as suggested by \cite{Witten:1996md}, while avoiding the introduction of a dynamical gauge field by introducing a new gauge symmetry that relates any gauge fields sharing the same topological data.

The `space of potentials' in the non-abelian  model for the three-form field considered in
\cite{Diaconescu:2003bm} consist of pairs $({\sf A}, {\sf c}_3)$, where ${\sf A}$ is a connection
on a bundle with gauge group $G$ and ${\sf c}_3$
is a `little' three-form. The group $G$ is taken to be $E_8$ in \cite{Diaconescu:2003bm} but we will also
consider other gauge groups. The combination of these which appears in various couplings is given by
\be
{\sf C}_3\ =\ {\sf c}_3 + CS({\sf A}) - \half CS(g)
\ee
where $g$ is the metric, and the field strength is given by
\be
{\sf G}_4\ =\ d{\sf c}_3 + {\rm Tr}({\sf F}\wedge {\sf F}) - \half {\rm Tr}({\sf R}\wedge {\sf R})
\ee
where ${\sf F}$ is the curvature of ${\sf A}$ and ${\sf R}$ is the Riemannian curvature associated to $g$.
The second and third term are supposed to capture the non-trivial topological part of the ${\sf G}$-flux.
In order for this data to reproduce the usual gauge equivalence classes for the three-form ${\sf C}_3$,
this data is subject to the gauge transformations (equation (3.14) in \cite{Diaconescu:2003bm})
\be
({\sf A}, {\sf c}_3) \ \to \ \ ({\sf A} + \psi, {\sf c}_3 - CS({\sf A}, {\sf A}+\psi) + \omega)
\ee
Here $\omega$, which is closed and has integral periods, generates the usual three-form gauge symmetry, whereas
$\psi$ is a non-abelian one-form gauge symmetry taking values in the same gauge group as ${\sf A}$. This one-form gauge symmetry locally gauges away all the degrees of freedom from the gauge field, so the bundle looks purely auxiliary. Indeed the usual Yang-Mills kinetic term is forbidden by the extended  gauge symmetry.

For later use we would like to point out that action on ${\sf A}$ of the one-form gauge symmetry generated by $\psi$ has actually been previously considered in topological gauge theory \cite{Brooks:1988jm,Baulieu:1988xs}. Accordingly we will often refer to it as a topological gauge symmetry.

Since ${\sf A}$ can be gauged away using the one-form symmetry, it is a fair question to ask what is achieved by introducing it in the first place. Indeed on a smooth and closed manifold it does not appear to be very interesting. But when the topological gauge symmetry is broken, we may get modes
that are not apparent from a `naive' formulation of the three-form.
For example in the presence of a boundary, it is natural to require $\psi$ to vanish at the boundary (or more precisely $\psi = d_A\lambda$ where $\lambda$ is a conventional zero-form gauge transformation). The associated edge modes comprise a non-abelian gauge field with only a conventional zero-form non-abelian gauge symmetry \cite{Diaconescu:2003bm,Baulieu:1988xs}.

We need to make a few comments on the choice of gauge group for the auxiliary topological gauge theory. In   \cite{Diaconescu:2003bm} the gauge group was taken to be $E_8$. The reason for this is the correspondence between classes ${\sf G} \in H^4(Y, {\bf Z})$ and isomorphism classes of $E_8$ bundles
on $Y$, with the mapping given by ${\rm Tr}({\sf F} \wedge {\sf F}) \to {\sf G}$. So one road we could take is to restrict ourselves to $E_8$, which is actually the relevant gauge group in the context of $F$-theory/heterotic duality. However the correspondence between $ H^4(Y, {\bf Z})$ and $E_8$ bundles is a statement about topological data and does not imply that any three-form can be represented as the Chern-Simons of an $E_8$ gauge field. Since we will use the Chern-Simons form explicitly in our argument for ADE singularities, this is perhaps one reason one would like to generalize the DFM model to other gauge groups for large rank ADE singularities. We will encounter other hints that the DFM model should be generalized, for example when considering the IIB limit of an $F$-theory compactification, where sheaves localized in complex codimension one seem to be the natural objects to consider.

\newsubsection{Compactification on ADE surfaces}
\subseclabel{ADE_surfaces}

We will be interested in the behaviour of the three-form in the presence of ADE singularities. So the simplest set-up would be for us to compactify $M$-theory on the corresponding ALE spaces. However from an algebro-geometric perspective, it is convenient to work with compactified versions of ALE spaces. Our focus will be on certain rational surfaces $S_{ADE}$ which were called ADE rational surfaces in \cite{LeungADE,LZ1,LZ2}. Roughly speaking,
these are obtained by taking an ALE space and adding a one-dimensional Calabi-Yau, i.e. an elliptic curve $E$, at infinity. For a precise description, see \cite{LeungADE,LZ1,LZ2}. Although not Calabi-Yau, the pair
$(S,E)$ is log Calabi-Yau, and admits a Calabi-Yau metric that diverges along $E$. We think of $E$ as the boundary of $S$. This set-up is a little richer than ALE spaces, but still rigidly constrained, and also appears naturally in string compactification. At any rate the results we are after have recently also been recovered for the ALE spaces themselves
\cite{LeungChenFlag}.

One of the central reasons for thinking about these ADE rational surfaces is that they provide
a natural way to study flat ADE bundles on the elliptic curve $E$.
Consider a log pair $(S_{ADE},E)$. The surface $S_{ADE}$ comes with a tautological ADE bundle ${\cal W}_S$, which by restriction to $E$ yields a flat ADE bundle ${\cal V}_E$ on $E$ (up to a twist). The main result of \cite{Friedman:1997ih,LeungADE,LZ1,LZ2} is that this
correspondence leads to an isomorphism of moduli spaces between pairs $(S_{ADE}, E)$ on one side, and flat ADE bundles on $E$ on the other side. The tautological bundle admits an intuitive expression. The cohomology lattice of the surface contains the corresponding ADE root lattice, so we may take a set of classes $\{\alpha\}$ that form an ADE root system. The tautological vector bundle associated to the adjoint representation can be expressed as
\be\label{W_adj}
{\cal W}_{\rm adj}  \ = \ \cO^{\oplus r} \oplus \sum_{\alpha\ \in\ {\rm roots}} \cO(\alpha)
\ee
The bundle is completely reducible, so the structure group is actually not the full ADE group $G_{ADE}$, but rather a maximal torus $T \subset G_{ADE}$.

This bundle has some very interesting properties. Perhaps most prominent for our purpose is the fact that $H^0({\cal W}_{\rm adj})$ and $H^1({\cal W}_{\rm adj})$ get extra generators precisely when we get effective curves in a class $\alpha \in {\rm roots}$, which signals that the bundle detects enhanced non-abelian gauge symmetry precisely when we expect it based on the wrapped $M2$-brane picture. Since the proof is straightforward, we briefly mention it here. By Serre duality $H^2(\cO(\alpha))\cong H^0(\cO(K - \alpha)) = 0$ because $K- \alpha$ cannot be effective (it has negative intersection with the proper transform of the hyperplane class of ${\bf P}^2$). Since $\chi(\cO(\alpha)) = 0$, we then have $\dim H^0(\cO(\alpha)) = \dim H^1(\cO(\alpha))$, but $\dim H^0(\cO(\alpha))> 0$ if and only if there is an effective curve with class $\alpha$, and then we will have $\dim H^0(\cO(\alpha)) = 1$ because the effective curve is a rigid $-2$-curve. 

We already know that ${\cal W}$ is the natural bundle to consider when restricted to the log boundary $E$ of $S_{ADE}$. We now want to argue that it is also the natural bundle to use in the context of the DFM model.

Suppose we compactify on a rational ADE surface or ALE space $S$. The usual story is that the three-form gives rise to $U(1)$ vector fields by expanding in a basis of harmonic two-forms $\omega^I \in H^{1,1}(S,{\bf C}) \cap H^2(S,{\bf Z})$:
\be
\label{KK_C3}
{\sf C}_3 \ =\ { A}_I \wedge \omega^I
\ee
The index $I$ may be thought of as labelling the Cartan generators $T^I \in {\bf g}$. It is convenient to take the basis $\{\omega^I \}$ to be given by 
$\{c_1(\cO(\alpha_i)) \}$ where $\alpha_i$ runs over the simple roots. The corresponding Cartan generators $T^{I_{\alpha_i}}$ are then part of a Chevalley basis for ${\bf g}$, i.e. they correspond to the coroots of the $\alpha_i$. The kinetic terms are proportional to
\be
\int_S  c_1(\cO(\alpha_i)) \wedge *\  c_1(\cO(\alpha_j)) = \vev{ \alpha_i , \alpha_j} = {\rm Tr}(T^{I_{\alpha_i}}T^{I_{\alpha_j}})
\ee
where $\vev{\alpha_i, \alpha_j} = - \alpha_i \cdot \alpha_j$ is the associated Cartan matrix, or equivalently minus the intersection product  on the surface.
An $M2$-brane wrapped on a cycle corresponding to a root $\beta$ would couple to the gauge field 
$A_{I_{\alpha_i}}$ as 
\be
\int_\beta c_1(\cO(\alpha_i)) \ =\ \alpha_i \cdot \beta
\ee
This is just what's expected from a root vector $T^\beta$ since
$[T^{I_{\alpha_i}}, T^\beta ] = \vev{\alpha_i ,\beta} T^\beta$ in a Chevalley basis.

Let us see what kind of bundle is compatible with this in the DFM model. First we claim that we may take the bundle to be holomorphic. One way to see this is as a consequence of the complex anti-self duality equations discussed in section \ref{TQFT}, by trivially extending our bundle on $S$ to a bundle on ${\bf C}^2 \times S$, which is a (non-compact) Calabi-Yau four-fold. The complex anti-self duality equations on ${\bf C}^2 \times S$ reduce to the equation ${\sf F}^{0,2}=0$ on $S$. Another way to see this is to use the gauge fixing using ordinary instanton equations for real four-dimensional manifolds in \cite{Brooks:1988jm,Baulieu:1988xs}. This leads to connections satisfying the real ASD equations on a four-manifold, which are equivalent to the Hermitian Yang-Mills equations on an algebraic surface.

We also expect a bundle which is completely reducible, i.e. a direct sum of line bundles, in order to match the generic $U(1)^r$ gauge symmetry in the lower dimensional theory. The bundle is then determined
by the first Chern classes of these component line bundles.

To get these Chern classes we can reason as follows. The Kaluza-Klein reduction of the three-form in (\ref{KK_C3}) involves non-trivial cohomology classes, so in the DFM model it would be natural for this part of the expansion to come from the Chern-Simons form of the bundle, so that upon further compactification we can naturally realize cohomologically non-trivial ${\sf G}$-fluxes as ${\rm Tr}({\sf F} \wedge {\sf F})$ of the bundle. Since the bundle on $S$ is abelian, we may write the relevant piece of the Chern-Simons form as ${\rm Tr}({ A} \wedge {\sf F})$, where ${\sf F}$ is meant to denote the curvature on $S$ (which has curvature only along the Cartan components) and $A$ the gauge field in the uncompactified dimensions. Expanding
\be
{\rm Tr}({ A} \wedge {\sf F}) \ = \ { A}_I \wedge {\rm Tr}(T^I {\sf F})
\ee
we see that the obvious way to get a match is to take the tautological bundle, which has
\be\label{tautological_T}
\omega^I \ =\  {\rm Tr}(T^I {\sf F})
\ee
where $T^I$ denote the Cartan generators. Indeed, expanding the curvature we get 
\be
 {\rm Tr}(T^{I_{\alpha_i}} {\sf F}) = \sum_j {\sf F}_{I_{\alpha_j}} {\rm Tr}(T^{I_{\alpha_i}} T^{I_{\alpha_j}}) = 
\sum_j {\sf F}_{I_{\alpha_j}} \vev{\alpha_i, \alpha_j}
\ee
But the curvature of the line bundle associated to a root vector $T^{\alpha_i}$ in (\ref{W_adj}) is given by $c_1(\cO(\alpha_i))$, in other words we have 
\be
[{\sf F}, T^{\alpha_i} ] \ = \ c_1(\cO(\alpha_i)) T^{\alpha_i}
\ee
Again expanding the curvature, and using the fact that the $\vev{\alpha_i , \alpha_j} $ also appear as structure constants in the Chevalley
basis (recall we had $[T^{I_{\alpha_i}}, T^\beta ] = \vev{\alpha_i , \beta} T^\beta$), we find that
\be
 {\rm Tr}(T^{I_{\alpha_i}} {\sf F}) = c_1(\cO(\alpha_i))
 \ee
in agreement with (\ref{tautological_T}). There are precisely $r$ equations for $r$ unknowns, where $r$ is the dimension of the Cartan of the ADE Lie group, so this fixes all the component line bundles. Thus, the Chern-Simons form of the tautological bundle captures the topological data of the three-form.

Instead of the identification in (\ref{tautological_T}), one seems to get further solutions by applying a non-trivial automorphism of the ADE root lattice. The inner automorphisms are given by Weyl permutations. Although this gives rise to different $T$-bundles (maximal torus bundles), these are the same as $G$-bundles. There are sometimes also outer automorphisms. These are in one-to-one correspondence with symmetries of the Dynkin diagram, so for $E_8$ for example there are none but for $SU(n)$ we get a $Z_2$ which maps a bundle to its dual (in other words, charge conjugation). However usually we are interested in the adjoint representation, and replacing the fundamental representation with its dual leaves the adjoint representation unaffected.

In the present paper we focus primarily on the complex structure, which is sufficient for $F$-theory. But we may also consider the $M$-theory picture. If we compactify $M$-theory on an ADE surface, the heterotic string doesn't just live on the log boundary of the surface, but rather on the real three-dimensional boundary at infinity, which is of the form $S^1 \times E$. The $S^1$ is the unit circle in the normal bundle to the log boundary $E$, so this is generally an  $S^1$-fibration rather than a direct product, but for simplicity we can consider the case of $dP_9$ where the normal bundle is trivial and we just get a direct product. In $M$-theory we also need to keep track of the  K\"ahler moduli of the surface, which are related to the complex structure moduli 
through a hyperk\"ahler structure. While the K\"ahler moduli do not affect the holomorphic structure of the tautological bundle, they do affect the hermitian metric 
and hence the connection on the bundle. Since the complex structure controls the holonomies of the tautological bundle on the elliptic curve $E$ at infinity, the hyperk\"ahler structure then dictates that the tautological bundle generally has non-trivial holonomies along the extra $S^1$ at infinity, and these holonomies are directly controlled by the K\"ahler moduli of the surface. This agrees with the heterotic string picture, which lives on
the $S^1\times E$ boundary at infinity and where we similarly get extra moduli from holonomies of the bundle along the extra $S^1$. In the $M$-theory picture, the point of enhanced gauge symmetry is achieved when the K\"ahler volumes of all the ADE exceptional cycles are set to zero. The hyperk\"ahler structure dictates that this sets the holonomies of the tautological bundle along the extra $S^1$ at infinity to zero, which agrees with the heterotic picture.

\newsubsection{Enhanced gauge symmetry}
\subseclabel{Enhanced_ADE}

We'd now like to take a closer look at the DFM data when the surface we are compactifying on develops an ADE singularity. Let us first describe
the general picture and then treat the case of an $A_1$ singularity more explicitly.

As we degenerate a smooth surface $S$ to get an ADE singularity, it turns out that the limit of the tautological bundle is not unique. It is also in general not a bundle (meaning the limiting sheaf is not locally free on the singular space), but since we are focussing on purely holomorphic
questions, it is convenient to resolve the singularity and work on the resolved surface. In physics this is often phrased as exploiting the decoupling of $F$-terms and $D$-terms. On the resolved surface, we do happen to get conventional vector bundles.

Intuitively, the tautological bundle is describing configurations of lines on the surface, whose cohomology classes correspond to the weights of some ADE representation, and on a smooth surface the bundle is completely determined. However when we get a singularity by colliding lines, there is some extra information in how exactly we collided the lines, and this gives rise to the different possible bundles in the singular limit. This extra information is needed for distinguishing different physical configurations associated 
to the same singularity. One can see a completely analogous picture on the log boundary, where one is effectively studying configurations of points, and one needs to keep track of the scheme structure as the points collide in order not to lose all the relevant information.

The different bundles appearing in the singular limit can be described explicitly in the following manner. 
There is a maximally reducible representative, which can again be written as 
\be
{\cal W}_{\rm adj}  \ = \ \cO^{\oplus r} \oplus \sum_{\alpha\ \in\ {\rm roots}} \cO(\alpha)
\ee
and where one
finds extra generators in its bundle-valued Dolbeault cohomology. We write the Kaluza-Klein expansion of the gauge field as
\be\label{KK_connection}
\delta {\sf A}^{0,1} \ = \ \Phi \wedge  \Xi^{0,1}_S + \delta A^{0,1} \wedge \Xi^{0,0}_S
\ee
where $\Phi$ and $A^{0,1}$ live in the uncompactified dimensions (${\bf R}^8$ in the case of $F$-theory), and $\Xi_S^{0,p}$ are representatives
of $H^p(S, {\cal W}_{\rm adj})$. As we discussed previously, for the completely reducible configuration
there is exactly one generator $\Xi^{0,1}_{S,\alpha} \in H^1(S, \cO(\alpha))$ and  one generator $\Xi^{0,0}_{S,\alpha} \in H^0(S, \cO(\alpha))$ 
whenever $\alpha$ is a positive root, as one 
would hope to find at the point of enhanced gauge symmetry. The negative roots come from the complex conjugates of these generators, or equivalently (as far as counting goes)
from the Serre duals. The deformations $\Xi^{0,1}_{S,\alpha} \in H^1(S, \cO(\alpha))$ are unobstructed and can be thought of as the missing non-abelian components of the adjoint field of the enhanced gauge symmetry at the singularity.\footnote{
In $M$-theory we expect an additional real adjoint scalar. It is guaranteed to be there by the hyperk\"ahler structure,
but as in the previous subsection it will depend on the hermitian metric and is not purely holomorphic.}
The other possible configurations in the singular limit are obtained by turning on a nilpotent expectation value
for the adjoint field, up to gauge transformations, i.e. they are parametrized by the nilpotent orbits of the corresponding ADE Lie algebra. 

The largest orbit
(also called the regular orbit) actually produces a bundle that is locally free
on the singular space; it corresponds to a nilpotent adjoint field that breaks the non-abelian gauge group down completely, and is obtained from the most generic collision of the lines. This bundle was discussed in some detail in \cite{FM,ChenLeung} as it is the only representative which actually corresponds to a vector
bundle on the singular surface, and hence perhaps deserves to be called `the' tautological bundle.
However it is evident that the representatives corresponding to the smaller orbits can arise in a less generic collision, i.e. by tuning additional parameters as we take the singular limit. These configurations are no less interesting and indeed the point of enhanced gauge symmetry (the maximally reducible representative)
is among them.

In this way the DFM data appears to be able to capture the degrees of freedom associated
to the enhanced non-abelian gauge symmetry, without ever referring to $M2$-branes wrapped on the vanishing cycles of the singularity. A similar pattern repeats itself for other singularities, like the conifold singularity.

Now of course as we have emphasized several times, deformations of the connection as in (\ref{KK_connection}) are normally gauge symmetries in the DFM model, so naively the DFM model cannot detect anything new in this manner. But just like when we have a boundary, when we have a singularity we should be careful about specifying what type of behaviour we allow for the gauge transformations. The crucial point to observe is that the generators $ \Xi^{0,1}_{S,\alpha}$ and $\Xi^{0,0}_{S,\alpha}$ have non-trivial behaviour at the exceptional locus and cannot be obtained
by pulling back smooth forms from the singular space to the resolved surface. We suspect that the harmonic representatives are in fact non-normalizable near the singularity for reasonable metrics, which is not allowed behaviour for gauge transformations. Thus we propose as a boundary condition that the one-form gauge transformations in the DFM model should remain smooth at the singularity. Then the extra deformations of the DFM model that appear at the singularity are not killed by the one-form gauge transformations and recover the missing non-abelian degrees of freedom of the ADE enhanced gauge symmetry.

The phenomena we have described can be understood from a local neighbourhood of the singularity, indeed they also hold for the ALE spaces (see section \ref{Grothendieck_resolution}). But as a consistency check, in the context of log del Pezzo surfaces we may also consider the asymptotic behaviour of the extra non-abelian modes that we found near the log boundary. We expect them to be non-normalizable at infinity, and in the case of $dP_9$ this actually follows from known results (see section \ref{log_CY}). Indeed they are directly related to the Dolbeault cohomology groups on the log boundary that are responsible for ADE enhancement in the heterotic string on the elliptic curve $E$. To see this,
the line bundle $\cO(\alpha)$ restricts to $\cO(p_1 - p_2)$ for some points $p_1,p_2$ on $E$. The cohomologies of $\cO(p_1 - p_2)$ are one-dimensional if $p_1 = p_2$ and zero otherwise. On $S$ the off-diagonal generators proportional to $T^\alpha$ are counted by $H^0(\cO(\alpha))$ and $H^1(\cO(\alpha))$, and these are non-zero and one-dimensional precisely when there is an effective curve with class $\alpha$. But then $\cO(\alpha)$ restricts to the trivial sheaf $\cO$ on $E$ and the Dolbeault cohomologies restrict to the corresponding generators 
of $H^0(\cO)$ and $H^1(\cO)$ on $E$, which is the statement of enhanced gauge symmetry for the Lie algebra generator $T^\alpha$ on the heterotic side.
As in \cite{Diaconescu:2003bm,Baulieu:1988xs} the natural condition is to restrict the one-form gauge transformations to preserve the bundle at infinity. So in this context we have an alternate way to see that these extra non-abelian modes can not be killed by the topological gauge transformations, because they also affect the bundle at infinity and the topological gauge transformations are not allowed to under our boundary condition (indeed there are no one-form gauge transformations shifting the gauge field in the heterotic string, it wouldn't even be compatible with the kinetic terms). 

For later use, let us expand a little bit on the above observation. In general if we have both a boundary and a singularity, then the natural boundary condition is to consider 
a generator of $H^1({\cal W}_{adj})$ a one-form gauge transformation if it restricts to zero both along the boundary and along the exceptional locus. However from the above remarks 
we conclude that for tautological bundles on rational surfaces, and presumably also for fibrations thereof, a generator of $H^1({\cal W}_{adj})$ that is non-vanishing along the exceptional locus is also non-vanishing along the boundary. Equivalently, if a generator of $H^1({\cal W}_{adj})$ vanishes along the boundary then it also vanishes along the exceptional locus. Hence for these particular geometries and bundles it is sufficient to require one-form gauge transformations in $H^1({\cal W}_{adj})$ to vanish along the boundary.

From the arguments above we recovered non-abelian gauge fields and adjoint scalars, but not the gaugino. When quantizing wrapped $M2$-branes, we are guaranteed to get a full supermultiplet and the vector multiplet is the only non-gravitational multiplet available in seven or eight dimensions with sixteen supercharges. Here too the configuration `looks supersymmetric' (as everything in sight looks holomorphic, even hyper-K\"ahler), which suggests the appearance of the gaugino may follow from the symmetries of the problem. Moreover from the remarks above, for del Pezzo surfaces with tautological bundles the same mechanism should produce the `heterotic' gaugino at the boundary at infinity. The original motivation for expressing the three-form as the Chern-Simons form of a bundle was in fact to allow the use of index theory in order to understand certain aspects of the $M$-theory partition function \cite{Witten:1996md}. This suggests there is some kind of higher dimensional adjoint fermion lurking in the background which gives rise to the physical gaugino in the heterotic string or at ADE singularities after imposing appropriate boundary conditions, something one may have suspected anyway given the anomaly inflow picture of \cite{Horava:1996ma}
(see \cite{Witten:2019bou} for a more familiar analogue of this).  At any rate, the origin of the gaugino remains to be clarified.

Thus we see that we can describe the phenomenon of non-abelian gauge symmetry enhancement by referring only to a consistent
description of the $M$-theory three-form field, without thinking about wrapped $M2$-branes. It would probably be interesting to think of other types of defects as well. For example we suspect that the non-abelian gauge symmetry appearing along $D$-branes can be understood as coming from a similar mechanism, which is supported by the fact that some $D$-branes originate as singularities in $M$- or $F$-theory.

To illustrate the mathematics in an explicit example, let us now consider the simplest
case of $SU(2)$ enhancement \cite{FriedmanMorgan,ChenLeung} which appears at an $A_1$ singularity. Again we work on the resolved version of the singular space.

For the simplest case of $SU(2)$ enhancement we have two distinguished lines $\ell_1$ and $\ell_2$ on the rational surface, 
with $\ell_1^2 = \ell_2^2 = -1$ and $\ell_1 \cdot \ell_2 = 0$. When the lines collide, we get an $A_1$ singularity, which we can resolve to get
an exceptional ${\bf P}^1$ in the class $\alpha = \ell_2 - \ell_1$ (or minus that), with $\alpha^2 = -2$. The line $\ell_2$ has effectively degenerated to the broken line $\ell_2 = \ell_1 + \alpha$. 

Now in the limit there are two bundles to consider, a reducible one and an irreducible one. Using the fundamental representation,
the reducible bundle is $\cO(\ell_1) \oplus \cO(\ell_2)$ which restricts to $\cO(1) \oplus \cO(-1)$ on the exceptional ${\bf P}^1$. Although this is a topologically trivial bundle on
${\bf P}^1$, it is not holomorphically trivial, and hence it does not correspond to a bundle on the singular space but to a sheaf. In particular the curvature is non-zero and when the ${\bf P}^1$ shrinks to zero size the connection will be not be smooth. As we've mentioned previously, since there is an effective curve of class $\alpha$ there is now a unique generator of ${\rm Hom}(\cO(\ell_1), \cO(\ell_2)) = H^0(\cO(\alpha))$  which means that the bundle has a non-abelian symmetry. Paired with this (through the index theorem) there is a single deformation in
$H^1(\cO(\alpha))$ which we want to interpret as the ``upper triangular'' non-Cartan generator of the adjoint field of the expected enhanced
$SU(2)$ gauge theory at the singularity. (The ``lower triangular'' generator is obtained as the Serre dual). It corresponds to deforming the bundle $\cO(1) \oplus \cO(-1)$ to the trivial rank two bundle $\cO(0) \oplus \cO(0)$ on the exceptional ${\bf P}^1$, and then lifting this to the rest of the rational surface. The resulting bundle ${\cal U}$ is just the unique non-trivial extension of $\cO(\ell_1)$ by $\cO(\ell_2)$:
\be\label{su2_extension}
0 \ \to \ \cO(\ell_2) \to \ {\cal U} \ \to \  \cO(\ell_1) \ \to \ 0
\ee
It is an irreducible rank two bundle and breaks the non-abelian $SU(2)$ symmetry. Since the deformation does not preserve the bundle along the exceptional ${\bf P}^1$,
it will be represented by a singular one-form in the limit that the ${\bf P}^1$ shrinks to zero, so if we want to think of this as a one-form gauge transformation then it is in fact a singular gauge transformation. Requiring one-form gauge transformations to be smooth at the singularity eliminates this deformation as a gauge transformation in the DFM model. More formally, 
generators of $H^1({\cal W}_{\rm adj})$ are allowed one-form gauge symmetries only if they are in the kernel of the natural restriction map to the exceptional locus.

To give a brief indication of how these objects arise as limits, we restrict to the boundary elliptic curve. The analogous story here has long been known an is explained for example in \cite{Friedman:1997ih}. The lines 
$\ell_1 $ and $\ell_2$ intersect the boundary elliptic curve in the points $p_1$ and $p_2$ respectively, so for smooth surfaces the tautological bundle restricts
to $\cO(p_1) \oplus \cO(p_2)$ on the boundary.  An extra twist is needed to turn this into a flat bundle, but this twist does not affect the following discussion.
Colliding the lines to generate an $A_1$ singularity corresponds to colliding the points to a single double point $p$ on the elliptic curve. 
Generically when we do that, on the boundary the limiting bundle is given ${\cal V}$ which is defined by the unique non-trivial extension
\be
0\ \to\ \cO(p)\ \to\ {\cal V}\ \to\ \cO(p)\ \to\  0
\ee
which we recognize as the regular representative (with the smallest automorphism group). This is just the restriction of the bundle ${\cal U}$ above to the boundary.
As we explained in  \cite{Donagi:2010pd}, from the Higgs bundle perspective it correspond to an adjoint field of the form 
\be
\Phi \ = \  
\left(
\begin{array}{cc}
0 & 1 \\
0 & 0 \\
\end{array}
\right)
\ee
i.e. it corresponds to the regular nilpotent orbit. But by tuning an additional parameter, we may also end up with the irregular representative $\cO(p) \oplus \cO(p)$, which is just the restriction of 
 $\cO(\ell_1) \oplus \cO(\ell_2)$ on the singular surface. 
 From the Higgs bundle perspective, it correspond to the adjoint field 
\be
\Phi \ = \  
\left(
\begin{array}{cc}
0 & 0 \\
0 & 0 \\
\end{array}
\right)
\ee

\newsubsection{Fibered version of the correspondence}
\subseclabel{Compactified_correspondence}

Due to the naturalness of the correspondence between rational surfaces and bundles on elliptic curves, we expect it should behave well in families. Thus by fibering over a common base $B$, we expect to get higher dimensional versions, analogous to the Higgs/spectral/ALE correspondence (for a review of this correspondence, see \cite{Wijnholt:2012fx}). We take $B$ to be a complex manifold of dimension $n$.
This leads to the following four classes of objects, which we claim are all related by some kind of transform, and are therefore all expected to be naturally equivalent:
\be\label{four_boxes}
\begin{array}{ccc}
  \fraam{(Z,{\cal V})} & \quad \longleftrightarrow \quad & \fraam{(Y,{\cal W})} \\[2mm]
  \updownarrow &  & \updownarrow \\[2mm]
  \fraam{(Z,{\cal L})} & \quad \longleftrightarrow \quad & \fraam{(Y,{\cal F})}
\end{array}
\ee
The data in these boxes consists of the following. $Z$ is an elliptic Calabi-Yau (with section)
of complex dimension $n+1$. We will take $Z$ to have a smooth Weierstrass model. We denote the fibration
by $\pi_Z:Z \to B$. ${\cal V}$ is a holomorphic $ADE$ bundle, semi-stable on the generic fibers of $\pi_Z$. The pair $(Z,{\cal V})$ is the analogue of the Higgs bundle. At some level we may think of it as a Higgs bundle on $B$
taking values
in elliptic curves \cite{DonagiCovers}.
The sheaf ${\cal L}$ is the spectral sheaf associated to ${\cal V}$, obtained by a
conventional Fourier-Mukai transform. Generically ${\cal L}$ is supported on a finite number of points on each of the fibers of $\pi_Z$. The pair $(Z,{\cal L})$ is clearly the analogue of the spectral data in the Higgs/spectral/ALE correspondence. This side of the correspondence is fairly well-known and established.

On the right hand side we have the analogue of the ALE fibrations. The manifold $Y$ is a complex $n+2$-fold, which is fibered over $B$
by ADE rational surfaces. We denote this fibration by $\pi_Y: Y \to B$. $Y$ may be constructed using the cameral cover, as explained in
section 3.1 of \cite{Curio:1998bva}, with a simplified version for the $A_n$ case using only the spectral cover discussed in \cite{ADE_Transform}. Our Calabi-Yau $n+1$-fold $Z$ is embedded in $Y$ as an anti-canonical divisor, so that the pair $(Y,Z)$ is a log Calabi-Yau,
and the fibrations $\pi_Y$ and $\pi_Z$ are compatible, in the sense that the elliptic fiber
$\pi_Z^{-1}p$ of a point $p \in B$ is the anti-canonical elliptic curve $E$ of the rational surface $\pi_Y^{-1}p$.
${\cal W}$ is a holomorphic bundle on $Y$, with structure group $G_{ADE} \times_{{\bf Z}_k} U(1)$, where ${\bf Z}_k$ is (a subgroup of) the center of $G_{ADE}$. (The extra $U(1)$ is not relevant for representations in which the center of $G_{ADE}$ acts trivially, such as the adjoint representation). The restriction of ${\cal W}$ to the fibers of $\pi_Y$ yields the tautological bundle associated to the fiber, and the restriction to the boundary $Z$ yields ${\cal W}|_Z = {\cal V}$. For the type $A_n$ case with smooth spectral cover and spectral line bundle, ${\cal W}$ was constructed in \cite{ADE_Transform} by a generalized Fourier-Mukai transform, and we expect our formula to hold more generally as long as the spectral sheaf is locally free. For the type
$E_n$ case, the correspondence between $(Z, {\cal V})$ and $(Y, {\cal W})$ was proven in \cite{Aker_dp_fibrations} when the base $B$ is a curve and under an analogous genericity assumption. Another version of the $E_n$ case is proven in \cite{UniTorsors_2016}.

Finally ${\cal F}$ is a sheaf supported on a certain divisor $R$ in $Y$, known as the cylinder,
i.e. the union of lines of each rational surface fibered over the base. The sheaf ${\cal F}$ may also be constructed by an integral transform, or equivalently by applying a version of the cylinder map:
\be
{\cal F}\  \sim\  i_{R*}p_R^*{\cal L}
\ee
The cylinder map is defined using the projection $p_R: R \to C$ from $R$ to the support $C$ of ${\cal L}$, and the embedding $i_R: R \to Y$. Again we expect this formula should hold (up to twisting) when ${\cal L}$ is locally free, and require only small modifications in some common cases where ${\cal L}$ is not locally free.
The restriction of ${\cal F}$ to the fibers of $\pi_Y$ yields the structure sheaf of the lines associated to that fiber (possibly up to some twisting), and the restriction to $Z$ yields ${\cal F}|_Z = {\cal L}$. Since ${\cal W}$ has been referred to as a tautological ADE bundle, we may refer to ${\cal F}$ as a tautological ADE sheaf.

In order to keep the present paper at least somewhat focused, we will work mostly with bundles and leave out a deeper discussion of the sheaf ${\cal F}$. But
we would like to mention here that the sheaf ${\cal F}$ seems to do just as well at capturing the physics of the $M$-theory three-form. For example in the situation considered in section \ref{Enhanced_ADE} where we had two lines $\ell_1$ and $\ell_2$ colliding to create an $A_1$ singularity, we again get two natural tautological sheaves, an irregular one given by $\cO_{\ell_1} \oplus \cO_{\ell_2}$ an a regular one given by $\cO_{\ell_1 + \ell_2}$, which can also be expressed by the unique extension analogous to (\ref{su2_extension}):
\be
0 \ \to \ \cO_{\ell_1} \ \to \ \cO_{\ell_1 + \ell_2} \ \to \ \cO_{\ell_2} \ \to \ 0
\ee
where $\ell_2 = \ell_1 + \alpha$. In fact, $F$-theory generically admits another limit, the Sen limit, in which the sheaf  ${\cal F}$ is the natural object to work with, as its restriction to the log boundary exactly yields the $D7$-branes of the perturbative type IIB theory \cite{Clingher:2012rg}. Thus it seems quite natural to believe that the DFM model should be generalized to the derived category, i.e. instead of
trying to relate the three-form to the Chern-Simons form of an $E_8$ bundle, we instead relate it to the Chern-Simons form associated to a whole complex of bundles or sheaves (i.e. the alternating sum of the Chern-Simons forms of each term in the sequence). As we noted previously, it seems unreasonable to expect to be able to express every three-form as the Chern-Simons form of an $E_8$ gauge field, and by going to the derived category one would expect to be able to recover a much larger set of three-forms as Chern-Simons forms, at the expense of introducing more gauge symmetries. This might perhaps even provide some kind of explanation for why derived categories appear in perturbative string theory in the first place.

Proving all the proposed correspondences (\ref{four_boxes}) would be a non-trivial technical undertaking, but morally speaking these are technical issues that arise from fibering the known and proven correspondence for surfaces. As mentioned we can prove special cases, and we can verify certain consequences predicted by the correspondence.

\newsubsection{${\sf G}_4$-flux from tautological bundles}
\subseclabel{bundle_to_flux}

Let us consider the ${\sf G}$-fluxes for the fibered geometries considered in section \ref{Compactified_correspondence}. 
Given that we found previously that the Chern-Simons form of a tautological bundle captures the 
topological information of the three-form. we would expect a simple relation
of the form ${\sf G}_4/2\pi  \sim {\rm ch}_2({\cal W}) - \half {\rm ch}_2(TY)$. On the other hand, in this set-up a conjectured expression for the ${\sf G}$-flux
for heterotic duals was previously given in \cite{Curio:1998bva} (with a small modification given in \cite{Hayashi:2008ba,Donagi:2008ca} to account for an extra blow-up that is needed). We will denote this latter $G$-flux by ${\sf G}_{CD}$. One's first thought might be that these expression have to agree, but this is actually not the case. We have ${\sf G}_{CD}|_Z = 0$, whereas ${\rm ch}_2({\cal W})|_Z = {\rm ch}_2(V)$ is non-zero, in fact explicit expressions for it may be found in \cite{Friedman:1997ih}.

So how do we decide what is the correct expression for the ${\sf G}$-flux dual to a heterotic model? The definition of the $\sf G$-flux in \cite{Curio:1998bva} was partly motivated by getting a match between tadpole cancellation in $F$-theory and the heterotic string. 
On the other hand an expression of the form ${\sf G}_4/2\pi  \sim  {\rm ch}_2({\cal W})$ is naturally suggested by the holomorphic Chern-Simons theory on the boundary $Z$, and ensures that the holomorphic Chern-Simons action agrees with the flux superpotential on $Y$ in $F$-theory. Indeed in section \ref{ADE_surfaces} we saw that upon compactification to eight dimensions the three-form matches with the Chern-Simons form of the tautological bundle, so the relation ${\sf G}_4/2\pi  \sim  {\rm ch}_2({\cal W})$ should just follow from fibering this correspondence over a base. Furthermore with the choice in \cite{Curio:1998bva} the flux superpotential on the four-fold seems to be well-defined, whereas the holomorphic Chern-Simons action for a bundle
$V$ on $Z$ only makes sense if ${\rm ch}_2(V)$ vanishes, again suggesting a problem with the definition in \cite{Curio:1998bva}.

There is however a simple explanation for this apparent discrepancy. The paper \cite{Curio:1998bva} had in mind gluing the log Calabi-Yau to another log Calabi-Yau to get a space $Y_1\cup_Z Y_2$ that is smoothable to a compact Calabi-Yau four-fold. In this situation, certain classes in $H^4(Y_1)$ and $H^4(Y_2)$ get identified when we map to $H^4(Y_1\cup_Z Y_2)$, and the dimension of $H^4$ will be further reduced when we go to a smoothing of the normal crossing singularities. Accordingly there will be cancellations between the ${\sf G}$-flux in $H^4(Y_1)$ and $H^4(Y_2)$. In particular the restriction from $Y_1$ to $Z$ would yield ${\rm Tr}({\sf F}_1 \wedge {\sf F}_1) - \half {\rm Tr}({\sf R} \wedge {\sf R})$ while the restriction from $Y_2$ to $Z$ would yield ${\rm Tr}({\sf F}_2 \wedge {\sf F}_2) - \half {\rm Tr}({\sf R} \wedge {\sf R})$ (now again including the half-integral shift), which cancels if tadpole cancellation is satisfied. Taking this into account and decomposing the cohomology as in in \cite{Curio:1998bva}, it appears that the projection to $H^2(B, \underline{\Lambda})$ (in the notation from \cite{Curio:1998bva}) is the only piece that needs checking. Furthermore we calculated this piece explicitly and it agrees exactly with ${\sf G}_{CD}$. In fact, the projection of  $ {\rm ch}_2({\cal F})$ also agrees with  ${\sf G}_{CD}$. We hope to discuss this calculation in more detail in \cite{F_Fluxes}. We have also studied matching the continuous deformations of the bundle with continuous deformations
of the three-form field, and again this works out under suitable smoothness conditions \cite{MF_Extension}.

In our setting here we prefer to work with the individual log-Calabi-Yaus without having an implicit gluing in mind, so it is more natural
for us to define the ${\sf G}$-flux as ${\rm ch}_2({\cal W}) - \half {\rm ch}_2(TY)$, rather than the piece that just lives in $ H^2(B, \underline{\Lambda})$.

As a corollary we get a much clearer conceptual picture of the mathematical content of heterotic/$F$-theory duality, in particular why the superpotentials on both sides should match: it basically boils down to the standard relation between topological gauge theory on a four-fold and Chern-Simons on its three-fold boundary.

\newsubsection{Stability walls}
\subseclabel{Stability}

One of the most notable features of the traditional formulation of $M$-theory or $F$-theory on Calabi-Yau
four-folds is the absence
of stability walls in K\"ahler moduli space \cite{Donagi:2010pd}. Indeed the $D$-terms are formulated as a primitiveness
condition on the ${\sf G}$-flux, viz. ${\sf J} \wedge {\sf G} = 0$, which is a closed condition and therefore fundamentally different from the stability conditions encountered in dual formulations. It's not hard to see that this discrepancy arises due to difficulties of the description of wrapped $M2$-brane states in the limit where these degrees of freedom are very light.

An interesting question therefore is whether the DFM model allows us to recover stability walls. Since we have now been able to recover all the
`semi-stable' configurations that the wrapped $M2$-brane picture didn't see, we should expect this to be possible.

For defining stability we generally need two ingredients. We need to be able to talk about sub-objects, and we need some kind of slope $\mu$. Then stability of an object $V$ is the requirement that $\mu(U) < \mu(V)$ for all sub-objects $U \subset V$.

In physics language one often rephrases this in terms of the Fayet model \cite{Kachru:1999vj}. Eg. if the gauge group for a principal bundle $P$ is some semi-simple Lie group, stability is the requirement that for any reduction of the structure group to a maximal parabolic subgroup, the slope is less than the slope of the original. One can associate a `relative' $U(1)$ subgroup to such a reduction (the center of the associated Levi) and an associated line bundle. The Fayet model is basically the effective field theory for this `relative' $U(1)$ gauge symmetry and the charged matter fields that describe deformations of the bundle to a reducible bundle. The slope of this relative line bundle is proportional to the Fayet-Iliopoulos term in the Fayet model, and stability amounts to checking the sign of the slope for this line bundle for every reduction to a maximal parabolic.
For $GL(n,{\bf C})$, reduction of the structure group to a  maximal parabolic subgroup is equivalent to writing the associated vector bundle $V$ as an extension
\be
0 \ \to \ U \to\ V\  \to\  V/U\  \to\  0
\ee
where both $U$ and $V/U$ cannot be further reduced in this way. The relative $U(1)$ is the structure group of the line bundle
\be
L \ =\  \det(U)^{1/r_1}\otimes \det(V/U)^{-1/r_2} 
\ee
where $r_1$ and $r_2$ are the ranks of $U$ and $V/U$ respectively. The charged matter fields in the Fayet model correspond to generators of $\Ext^1(V/U, U) $. The Fayet-Iliopoulos parameter $\xi$ is the slope of the relative $U(1)$ line bundle $L$, which is ${\rm deg}(U)/r_1 - {\rm deg}(V/U)/r_2$. It is easy to show that $\xi$ is negative iff ${\rm deg}(U)/r_1 < {\rm deg}(V)/(r_1 + r_2)$. 

Since the DFM model uses a non-abelian bundle in its formulation, we can use the same notion of checking reductions of the structure group to a maximal parabolic
in the notion of stability. However there are some modifications of the familiar story. First, we are interested only in reductions that are physically distinct, i.e. we need to identify configurations that are related by valid one-form gauge symmetries. Second, since we don't have conventional propagating gauge fields, we get a notion of moment maps or Fayet-Iliopoulos parameters for $U(1)$ gauge symmetries that differs from the slope.
Let us denote by $\omega_Y$ the two-form in $H^2(Y_4)$ associated to the relative $U(1)$ gauge symmetry, which we can also interpret as $\omega_Y = c_1(L_Y)$ where $L_Y$ is the line bundle associated to the relative $U(1)$. Then $M/F$-theory tells us not to use the slope of $L_Y$ for defining stability, which would be proportional to the degree $\int {\sf J}^3 \wedge \omega_Y$, but rather use the $M/F$-theory expression \cite{Haack:2002tu,Donagi:2008kj}
\be
\xi_Y \ \propto \ \int_{Y_4} {\sf G} \wedge {\sf J} \wedge \omega_Y
\ee
In the set-up of $F$-theory/heterotic duality one sees that it reproduces the standard heterotic criterion of slope stability and stability walls in the appropriate limit \cite{F_Stability}, at least for structure groups of type $A_n$. Morally that's because the correspondence between the heterotic and the $F$-theory side is given by a type of Fourier-Mukai type transform \cite{ADE_Transform}, so then we just need to check that the Fayet-Iliopoulos parameters on both sides agree in the appropriate limit, which happily they do.

\newsubsection{ALE spaces and the Grothendieck-Springer resolution}
\subseclabel{Grothendieck_resolution}

Coming back to the Higgs/spectral/ALE correspondence, in view of the above
we expect that the ALE side should be generalized in the same way
in order to get a complete mathematical equivalence between the different pictures. 
In other words, it is necessary to keep track of some extra information on the ALE surface, which 
we can do by considering the pair $(S, {\cal W})$ where $S$ is the ALE surface and 
${\cal W}$ a tautological bundle. Since (after resolving possible singularities)
the Picard lattice of the ALE can be identified with the corresponding root lattice, we can write
our usual tautological bundle
\be\label{taut_bundle}
{\cal W}  \ = \ \cO^{\oplus r} \oplus \sum_{\alpha\ \in\ {\rm roots}} \cO(\alpha)
\ee
though as seen previously on a singular surface we will also get some irreducible bundles related by deformation. 
As in section \ref{Enhanced_ADE}, we expect that the deformation theory of an ADE singularity and a tautological bundle on it is controlled by a complex scalar
field $\Phi$ in the adjoint representation, i.e. a copy of the ADE Lie algebra ${\bf g}$ modulo the adjoint action of the corresponding Lie group $G$. (Since we ignore $D$-terms here, we always work with the complexified versions). 

The geometry of the adjoint quotient, and in particular the Grothendieck-Springer resolution, allows one to unify various different classification problems in which ADE Dynkin diagrams appear into a single mathematical framework. It is therefore a natural question if the physicists' prediction of ADE enhanced gauge symmetry can also be related to this framework, and as we will see the results of reference  \cite{LeungChenFlag} together with their interpretation in the present paper provides an answer to this question. In the process we will also see the relation between tautological bundles on the ALE and Bott-Borel-Weil line bundles. Let us briefly review the set-up. 

For an ADE Lie algebra ${\bf g}$ we consider the adjoint quotient map
$\psi: {\bf g} \to {\bf t}/W$ which sends an element $g\in {\bf g}$ to its invariants.
Here $W$ denotes the associated Weyl group. The fiber $\psi^{-1}(0)$ over $0 \in {\bf t}/W$
is the nilpotent cone $N({\bf g})$ and the other fibers are deformations of this singular variety. The fibers are holomorphic symplectic varieties, in particular they are Calabi-Yau.
The Grothendieck resolution
$\tilde{\tilde{\bf g}} \to {\bf g}$ is the simultaneous resolution of this family. Explicitly it consists of pairs
$(g, {\bf b})$ where $g\in {\bf g}$ and ${\bf b}$ is a Borel subalgebra containing $g$. We can also write it as
$\tilde{\tilde{\bf g}} = (G \times {\sf b})/B$ where ${\bf b}$ is a fixed Borel subalgebra, and $B$ is the corresponding
Borel subgroup. The map $\tilde{\tilde{\bf g}} \to {\bf g}$ is given by $(g,x) \to g x g^{-1}$. We can fit the Grothendieck resolution in a diagram
\be
\begin{array}{rcccl}
& \tilde{\tilde{\bf g}} & \to&  {\bf g}& \\[1.5mm]
 &\!\!\!\tilde{\tilde{\psi}} \downarrow & & \,\,\downarrow \psi & \\[1.5mm]
& {\bf t} & \to & {\bf t}/W &
\end{array}
\ee
The fiber $\tilde{\tilde \psi}^{-1}(0)$ gives the resolution of the nilpotent cone  $(G \times {\sf n})/B \to N({\bf g})$, also known as the Springer resolution. The Springer resolution is isomorphic to $T^*(G/B)$, the cotangent bundle over the flag manifold $G/B$.

For convenience we also introduce the intermediate space $\tilde {\bf g}$. It is obtained by pulling back the covering ${\bf t} \to {\bf t}/W$ using $\psi$. Although ${\bf g}$ is just a vector space and thus not singular, the variety $\tilde {\bf g}$ is singular.  We have a map $\tilde {\tilde {\bf g}} \to \tilde {\bf g}$ which resolves the singularities.

The connection with ADE singularities comes about because the closures of nilpotent orbits often have singularities of the corresponding type. 
For example, the closure of the maximal nilpotent orbit for $SL(2)$ is given by an equation $xy-z^2=0$ which we recognize as an $A_1$ singularity. 
To formalize this, we make some definitions.
An element $g \in {\bf g}$ is said to be regular if the dimension of its orbit is maximal, equivalently if the dimension
of its stabilizer is minimal. An element is said to be subregular if the dimension of its orbit is two less than maximal.
For a subregular element $x$, we may consider a slice $S_x$ in ${\bf g}$ transversal to the orbit through $x$. This is called a Slodowy slice. The intersection with the central fiber $S_x \cap \psi^{-1}(0)$ yields the ALE space ${\bf C}^2/\Gamma_{ADE}$, and $S_x$ itself describes the whole semi-versal family of deformations of ${\bf C}^2/\Gamma_{ADE}$.
By lifting to the Grothendieck resolution, we get the simultaneous resolution of the whole deformation family of an ALE space.

Using this picture we can build natural tautological bundles on the Grothendieck resolution $\tilde{\tilde{\bf g}} $, and by restriction to the Slodowy slice $S_x$ we then get natural tautological bundles on the whole deformation family of an ALE space.
Let us start with the flag manifold $G/B$, where the degree two cohomology
$H^2(G/B)$ is given by the weight lattice. This contains the root lattice, so we can write a reducible tautological Lie algebra bundle on $G/B$, which is just a sum of Bott-Borel-Weil line bundles associated to the roots, then pull it back to $\tilde{\tilde{\bf g}}$ using the observation that $\tilde{\tilde{\bf g}} = (G \times {\sf b})/B$.

Reference \cite{LeungChenFlag} studied the pull-back of the tautological bundle to $(G \times {\sf n})/B$, which corresponds to the Springer resolution (the central fiber of the map $\tilde{\tilde{\bf g}} \to {\bf t}$). The intersection with the subregular Slodowy slice yields the resolution of ${\bf C}^2/\Gamma_{ADE}$, and this construction does indeed yield the expected tautological bundle (\ref{taut_bundle}) on the resolved ALE. Let $\alpha$ correspond to a root.
To get the non-abelian
enhancement, as for the rational ADE surfaces we would to see that $H^0(\cO(\alpha))$ and $H^1(\cO(\alpha))$ are one-dimensional on the ALE surface precisely when there is an effective curve in the class $\alpha$. According to \cite{LeungChenFlag}
this holds for degree one and when $\alpha$ is a simple root. It is also immediate for degree zero. The remaining degree one Dolbeault cohomology groups were not computed but the result follows from the symmetries of the problem (by taking commutators of the $H^0$ and the $H^1$ Dolbeault cohomologies). 

One can explicitly follow the nilpotent deformations of the non-abelian adjoint fields of the gauge theory; these correspond to the bundle deformations on the resolved ALE studied in \cite{LeungChenFlag} analogous to the deformations encountered in section \ref{Enhanced_ADE}, and are simply inherited from the bundle deformations on the Springer resolution that deform the reducible Lie algebra bundle considered above to the holomorphically trivial Lie algebra bundle $(G \times {\bf n} \times {\bf g})/B \to (G \times {\bf n})/B$. These relations with geometric representation theory suggest that the phenomenon of enhanced gauge symmetry at singularities is closely tied with the symmetries of the problem, and perhaps a consequence of it.

\newsubsection{Frozen singularities}

Not long after the phenomenon of enhanced gauge symmetry at ADE singularities was discovered, it was noticed that the $D$ and $E$-type singularities also appear to have `frozen' variants 
with reduced gauge symmetry \cite{Witten:1997bs,deBoer:2001wca}. Compared to the usual versions, they appear to support a discrete three-form flux. In the heterotic description,
one finds corresponding bundles that are not completely reducible. For example for the $Spin(32)/Z_2$ case \cite{Witten:1997bs},
on the heterotic side one needs a bundle `without vector structure' which at a generic point in moduli space is a sum of irreducible stable $PGL(2)$-bundles of degree one. Therefore in the DFM model, we should similarly expect that frozen ADE singularities are distinguished from ordinary singularities by virtue of the three-form being associated to a bundle that is not completely reducible. The fact that bundles can obstruct deformations is of course a well known phenomenon which we will meet again in section \ref{Bulk_boundary_compare}.

Here we would like to make some brief comments on the possible construction of bundles relevant for the frozen version of the $D_n$ singularity. We have not checked details however, so these comments should be considered as very preliminary.

The frozen version of the $D_n$ singularity is known to involve $Sp$ gauge groups, indeed it gives rise to $O^+$-planes in perturbative string theory \cite{Witten:1997bs}. The $D_n$ type ALE singularities can be realized as $Z_2$ quotients of $A_{2n-1}$ singularities. The Dynkin diagram for $Sp(n)$ can be obtained by folding the $A_{2n-1}$ Dynkin diagram by a $Z_2$ outer automorphism. This suggest we can construct a natural $Sp(n)$ bundle on the $D_n$ singularity by quotienting the tautological $A_{2n-1}$ bundle
by an outer automorphism (i.e. we consider a sub-bundle where the $Z_2$ identification is twisted by an outer automorphism). The Chern-Simons form for this bundle would be the natural candidate for the three-form.

For the $E_8$ case, it was pointed out in \cite{deBoer:2001wca} that the data of an $E_8$ triple is equivalent to the data of a rational elliptic surface with an automorphism. The quotient of the surface by the automorphism yields another rational elliptic surface, so it again seems very tempting to take tautological bundles upstairs and twist the quotient bundle by the corresponding automorphism.

\newpage

\newsection{Holomorphic TQFT}
\seclabel{TQFT}

\newsubsection{Superpotential}

The superpotential on a Calabi-Yau four-fold $Y$ with flux ${\sf G}_4$ is given by \cite{Becker:1996gj,Gukov:1999ya}
\be\label{G_super}
W = \int_Y \Omega^{4,0}\wedge {\sf G}_4
\ee
As we reviewed in section \ref{DFM_model}, the DFM model for the three-form \cite{Diaconescu:2003bm} consists of pairs $({\sf A}, {\sf c}_3)$.
with curvature
\be
{\sf G}_4  \ =\ d{\sf c}_3 + {\rm Tr}({\sf F} \wedge {\sf F})-\half  {\rm Tr}({\sf R} \wedge {\sf R})
\ee
so substituting in (\ref{G_super}) yields the superpotential for the data $({\sf A},{\sf c}_3)$.

The model has a topological gauge symmetry ${\sf A} \to {\sf A} + \psi$ which kills all the local degrees of freedom of the gauge field.
In the following, we want to gauge fix this topological gauge symmetry. To this end we will focus on the gauge sector of the superpotential, i.e. we take
\be\label{p1_super}
W  \ = \
\int_Y \Omega^{4,0} \wedge  {\rm Tr}({\sf F} \wedge {\sf F}) 
\ee
and translate back to the three-form field when needed. A convenient gauge fixing has been carried out in  \cite{Baulieu:1997jx}, and we will see
many of the expected features arise. After this gauge fixing, the infinite dimensional set of one-form gauge transformations of the DFM model is reduced to a finite dimensional set of gauge transformations, described by Dolbeault cohomology groups of the bundle.

A more complete treatment would perhaps treat the gauge symmetries of ${\sf A}$ and ${\sf c}$ jointly. A model along these lines
was actually also considered in \cite{Baulieu:1997jx}, but the gauge transformations there act slightly differently. We will not explore this further here.

\newsubsection{Holomorphic version of topological Yang-Mills theory}
\subseclabel{Holomorphic_DW}

The gauge theory with action (\ref{p1_super}) and gauge symmetry ${\sf A} \to {\sf A} + \psi$ on a Calabi-Yau four-fold
is essentially the complex analogue of topological Yang-Mills theory on ordinary differential four-manifolds \cite{Witten:1988ze,Baulieu:1988xs}. In the complex version we replace the real four-manifold with a Calabi-Yau four-fold, $i$-forms by $(0,i)$-forms, the Hodge $*$-operator by
$\star = \bar{\ast}(\cdot \wedge \Omega^{4,0})$, and the quadratic intersection pairing on two-forms by the Serre duality pairing
\be
Q(\alpha, \beta) \ = \ \int_Y \Omega^{4,0}\wedge {\rm Tr}(\alpha \wedge \beta)
\ee
(The analogue of a boundary is the notion of log-boundary, which we will encounter as well). In the complex case
the gauge fixing of the topological symmetry was carried out in \cite{Baulieu:1997jx}. Just as we ended up with the twisted $4d$ $N=2$ Yang-Mills theory
in the real four-dimensional case, so we end up with the twisted $8d$ super-Yang-Mills theory in the Calabi-Yau four-fold case.

Let us briefly review some aspects of the $8d$ supersymmetric Yang-Mills theory with gauge group $G$. It is simply obtained by dimensional reduction from the
ten-dimensional super-Yang-Mills theory with gauge group $G$, with action
\be
\int -{1\over 4 g^2} {\sf F}^{\mu\nu}{\sf F}_{\mu\nu} + {i\over 2 g^2} \lambda \not\! D_{\sf A} \lambda
\ee
where in the above equation ${\sf A}$ denotes the ten-dimensional gauge field and $\lambda$ is the ten-dimensional gaugino.
After reduction, the field content is as follows. The bosonic fields consist of an eight-dimensional vector field ${\sf A}$ and two 
real scalars in the adjoint representation of $G$, which
we can combine into a complex scalar $\phi$. There are two real gauginos of positive and negative chirality. On a K\"ahler four-fold we may identify the complexified spinor bundles as $S^+ \otimes {\bf C} = \Lambda^{0,{\rm even}}$ and
$S^- \otimes {\bf C} = \Lambda^{0,{\rm odd}}$. Real spinors are fixed points of a conjugation which we may identify with the $\star$ operation.. Identifying $\Lambda^{0,0}$ with the fixed points in $\Lambda^{0,0} \oplus \Lambda^{0,4}$ and
$\Lambda^{0,1}$ with the fixed points in $\Lambda^{0,1} \oplus \Lambda^{0,3}$, we are left with
\ba\label{CY4_spinors}
S^+\! \otimes\! Ad(G) \is \Lambda^{0,0}(Ad(G)) \oplus \Lambda^{0,2}_+(Ad(G)) \eol[2mm]
S^-\! \otimes\! Ad(G) \is \Lambda^{0,1}(Ad(G))
\ea
We denote their sections as $\eta^{0,0}, \chi^{0,2}_+$ and $\psi^{0,1}$ respectively. It is sometimes convenient to have bosonic auxiliary fields
$B^{0,2}_+$ which are identified with ${\sf F}^{0,2}_+$ by the equations of motion and which can be integrated out.

The BPS equations are obtained by dimensional reduction of the usual expression
\be
\delta \lambda \ = \ {\sf F}_{\mu\nu} \gamma^{\mu\nu}\zeta \ = \ 0
\ee
where $\zeta$ is a covariantly constant spinor, and $\gamma^{\mu\nu} = {1\over 4} [\gamma^\mu, \gamma^\nu ]$ is the usual anti-commutator constructed from the Clifford matrices. When the adjoint scalars vanish, one may verify \cite{Acharya:1997jn,Bak:2002aq} that this is equivalent
to the `octonionic' instanton equations in eight dimensions:
\be
{\sf F}_{\mu\nu} + {1\over 2}{\sf T}_{\mu\nu\kappa\lambda}\, {\sf F}^{\kappa \lambda} \ = \ 0
\ee
where ${\sf T}^{\mu\nu\kappa\lambda} = \zeta^T \gamma^{\mu\nu\kappa\lambda}\zeta$ is the invariant rank four tensor of $Spin(7)$ constructed from the octonions. On a manifold of $SU(4)$ holonomy we have
${\sf T} \sim \half {\sf J} \wedge {\sf J} + {\rm Re}(\Omega^{4,0})$ and these equations fall apart into
\be
{\sf F}^{0,2}_+ \ = \ 0, \qquad {\sf F} \wedge {\sf J}^3 \ = \ 0
\ee
where ${\sf F}^{0,2}_+$ is the self-dual part of ${\sf F}^{0,2}$ with respect to $\star$. With the spinor bundles as in (\ref{CY4_spinors}), the Dirac operator may be identified with
\be\label{Dirac_op}
\delb^\dagger \oplus \delb_+ :\ S^-\!\otimes\! Ad(G) \to S^+\! \otimes\! Ad(G).
\ee
The fermionic zero
modes are then exactly the harmonic forms
of the elliptic complex
\be
0\ \to \ \Lambda^{0,0}(Ad(G)) \ \to^{\delb}\ \Lambda^{0,1}(Ad(G)) \to^{\delb_+} \Lambda^{0,2}_+(Ad(G)) \ \to \ 0
\ee
This is wholly as expected because the above complex describes linearized deformations of the
complex ASD equations modulo gauge symmetries. All this fits nicely with the picture advocated in \cite{Donaldson:1996kp,ThomasGaugeTheory} regarding the complex analogues of well-known topological field theories.

The $D$-term equation ${\sf F} \wedge {\sf J}^3=0$ can be formulated as slope stability of the bundle. While this would be the natural condition say when compactifying the heterotic string on a four-fold, in $M$- and $F$-theory we seem to be naturally led to a different stability condition. In the present paper however we restrict ourselves to purely holomorphic questions. In this case the precise stability condition is not crucial.

\newsubsection{Reduction}

While our main focus here is on Calabi-Yau four-folds, let us briefly recall some other configurations encountered in the `gauge theory in higher dimensions' program and see how they might be used for the study of $M$-theory compactifications. 

As we briefly mentioned above, the gauge fixing procedure for the one-form topological gauge symmetry ${\sf A} \to {\sf A} + \psi$ in \cite{Baulieu:1997jx} can also be performed on a ${\rm Spin}(7)$-manifold using the ${\rm Spin}(7)$ instanton equations, which we rewrite here as:
\be
{\sf T}_4 \wedge {\sf F}\  =\ - *_8 {\sf F}
\ee
where ${\sf T}_4$ is the harmonic Cayley four-form associated to the ${\rm Spin}(7)$ structure.  
So we could study $M$-theory on a ${\rm Spin}(7)$ manifold $X_8$ by relating the $M$-theory three-form to the Chern-Simons form of ${\rm Spin}(7)$ instantons. 

By dimensional reduction, we obtain various other related equations.
For example if we take a ${\rm Spin}(7)$ manifold of the form $X_8 = X_7 \times {\bf R}$ where $X_7$ is a $G_2$ holonomy manifold, and we consider bundles on $X_8$ that are pulled back from $X_7$,
then the ${\rm Spin}(7)$ instanton equations reduce to the $G_2$-instanton equations on $X_7$:
\be
\Psi_3 \wedge {\sf F}  = - *_7 {\sf F}
\ee
where $\Psi_3$ is the harmonic three-form associated to the ${\rm G}_2$-structure.
So we can study $M$-theory on $G_2$ manifolds by relating the $M$-theory three-form to the Chern-Simons form of a $G_2$-instanton bundle. One would hope that this similarly gives more insight into $M$-theory on ${\rm G}_2$-manifolds with defects such as boundaries and singularities.
In particular, by analogy with what we have seen in the present paper, we expect that the Higgs bundle/spectral cover/ALE-fibration correspondence for ALE-fibered $G_2$-manifolds \cite{Pantev:2009de} is most naturally formulated by putting such $G_2$-instanton bundles on the ALE-fibered manifold. One would hope such bundles can be constructed by putting tautological bundles on the ALE and using a real analogue of the Fourier-Mukai transform in \cite{ADE_Transform}. 

One may also consider the situation where the $G_2$-manifold has a Calabi-Yau three-fold boundary. Then the restriction of the $G_2$-instanton bundle should yield the heterotic bundle expected from $M$-theory/heterotic duality. This is a  real analogue of the relation between $F$-theory on a log Calabi-Yau four-fold and the heterotic string on its Calabi-Yau three-fold boundary considered in this paper.

By an analogous reduction to $6d$ we get Hermitian Yang-Mills. So for $M/F$-theory on a Calabi-Yau three-folds $X_6$, we would again naturally relate the three-form to holomorphic bundles on $X_6$, etc.

\newsubsection{Local structure of the moduli space}
\subseclabel{Local_moduli}

This is all very interesting conceptually, but in practice we would rather deal with holomorphic bundles and sheaves than with differential-geometric ASD solutions, because then
we can use techniques from algebraic geometry. Naively these objects are rather different. The hermitian Yang-Mills equations on a four-fold involve $(6 + 6 + 1)\times d$ equations (vanishing of the self-dual and anti-self-dual parts of ${\sf F}^{(0,2)}$, and vanishing of the K\"ahler trace of ${\sf F}^{(1,1)}$, and we have $d$ gauge invariances for the $8\times d$ unknowns of a vector field, and are thus overdetermined. Here $d$ denotes the dimension of the gauge group (eg. $d = n^2$ for $U(n)$). On the other hand the complex ASD equations (supplemented by the standard moment map equation) involve $(6 + 1)\times d$ equations and $d$ gauge invariances for $8\times d$ unknowns, and are thus `precisely right' in eight dimensions. Nevertheless there is a close relation between the solutions of these equations and there has been significant recent progress comparing the two \cite{Brav:2013,Cao:2014bca,Borisov:2015vha}.

We start with an observation from \cite{Lewis:1998}.
By decomposing ${\sf F}^{0,2}$ into its self-dual and anti-self-dual parts, we see that
the superpotential may be decomposed as
\be\label{SD_diff}
 \int_Y \Omega^{4,0} \wedge {\rm Tr}({\sf F}^{0,2} \wedge {\sf F}^{0,2}) \ = \
 || {\sf F}^{0,2}_+||^2 \ - \  ||{\sf F}^{0,2}_-||^2
\ee
But the left hand side is invariant under a continuous deformation of the gauge field. It follows that if the bundle
admits a holomorphic connection, so that the left hand side is zero, then solutions of the ASD equation ${\sf F}^{0,2}_+ = 0$ on the same connected component are also solutions
of ${\sf F}^{0,2} = 0$, even though the latter involves twice as many equations. In other words if such a holomorphic connection exists, then the moduli spaces ${\cal M}_{ASD} = {\cal M}_{\rm hol}$ of ASD instantons and holomorphic bundles are equal as sets. This is the holomorphic analogue of a similar relation between flat and ASD connections in four dimensions.
However even though ${\cal M}_{ASD}$ and ${\cal M}_{\rm hol}$ are equal as topological spaces, due to the fact that
the complex ASD equations are only a subset of the Hermitian Yang-Mills equations, the two spaces should differ in their
non-reduced structures.

To proceed further, instead of the infinite dimensional description of ASD connections modulo gauge equivalence, we set up a finite dimensional local model for the moduli space. In physics terms, such a local model usually arises from `integrating out the massive modes,' although in the present context that seems somewhat less clear due to the topological gauge symmetry floating around. We parametrize small deformations $\delta {\sf A}^{0,1}$ in terms of tangent directions to the moduli space, which are generators of $\Ext^1({\cal W},{\cal W})$. Physically, these would be the massless modes. Under such a deformation we roughly speaking have ${\sf F}^{0,2}\to {\sf F}^{0,2} + \kappa(\delta{\sf A}^{0,1})$, where
$\kappa: \Ext^1({\cal W},{\cal W}) \to \Ext^2({\cal W},{\cal W})$. We review this in a bit more detail in appendix \ref{Deformation_Theory}. If we view the classes
of $\delta {\sf A}^{0,1}$ as defining local coordinates on the moduli space, then $\kappa$ can be viewed as a section of a bundle $E$ over our local patch of the moduli space whose fibers have the same dimension as the vector space $\Ext^2({\cal W},{\cal W})$. The quadratic form $Q$ on $\Ext^2({\cal W},{\cal W})$ induces a quadratic form on $E$ which we denote by the same name, and due to deformation invariance of (\ref{SD_diff}) if ${\sf F}^{0,2}$ vanishes then we might expect to be able to choose $\kappa$ such that $Q(\kappa,\kappa) = 0$. This has been investigated in \cite{Brav:2013} and indeed they showed that moduli spaces of bundles or sheaves
on Calabi-Yau four-folds can be locally described by the following data:
\begin{itemize}
  \item a small neighborhood $U$ of 0 in $\Ext^1({\cal W},{\cal W})$;
  \item a vector bundle $ E\to U$, whose fibers have dimension given by $\Ext^2({\cal W},{\cal W})$;
  \item a non-degenerate quadratic form $Q(\cdot,\cdot)$ on $E$;
  \item a holomorphic section $\kappa \in H^0(E)$ (a Kuranishi map), such that $Q(\kappa,\kappa) = 0$.
\end{itemize}
The moduli space is then locally described as $\kappa^{-1}(0)$. The fact that $\kappa$ can be taken to satisfy
$Q(\kappa,\kappa)=0$ is to be contrasted
with Calabi-Yau three-folds, where there is no such structure.

Furthermore, the above data can be packaged in a holomorphic potential function $\Phi_{BV}$. It is defined on the space
of all the Ext groups $\Ext^*({\cal W},{\cal W})$, which are interpreted as the fields and anti-fields in the Batalin-Vilkovisky formalism.\footnote{
Roughly speaking, using the Batalin-Vilkovisky formalism to describe degenerate critical points of action functionals is what algebraic geometers refer to as derived geometry.
See eg. \cite{Stasheff:1997iz,Costello:2016vjw} for explanations.}
Restricted to generators of degree one (denoted by $\delta {\sf A}^{0,1}$) and degree
two (denoted by $B^{0,2}$), this potential takes the form
\be
\Phi_{BV} \ = \ Q(\kappa(\delta{\sf A}^{0,1}), B^{0,2})
\ee
It satisfies a classical master equation $\{ \Phi_{BV},\Phi_{BV}\} = 0$, which amounts to the condition
$Q(\kappa,\kappa)=0$. As we will discuss further in section \ref{Obstructions}, the restriction of $\Phi_{BV}$ to the physical fields
is a natural candidate for the effective superpotential as a function of the massless modes.

To see the relation with complex ASD solutions, we may split $E$ in to maximal sub-bundles
$E = E_+ \oplus E_-$ on which $Q(\cdot,\cdot)$ is positive or negative definite, and write $\kappa = (\kappa_+, \kappa_-)$.
Similar to (\ref{SD_diff}), the equation $Q(\kappa,\kappa)=0$ then implies that $| \kappa_+|^2 = |\kappa_-|^2$, and hence
$\kappa^{-1}(0)$ and $\kappa^{-1}_+(0)$ are identical as sets. But $\kappa^{-1}_+(0)$ locally describes the space of solutions
of the
complex ASD conditions. These local reductions can also be glued together globally according to
the results of \cite{Borisov:2015vha}, but this is not needed in our setting.

So we see that while we recover the same set of solutions (provided there exists at least one hermitian Yang-Mills connection on the same connected component), the Hermitian Yang-Mills equations yield too many
equations for this set. Roughly speaking the space of solutions of the hermitian Yang-Mills equations may locally be given by $x^2 = 0$ whereas the corresponding space of solutions of the complex ASD equations may be given by $x = 0$. There is a holomorphic map between them by 'killing' some equations (forgetting the nilpotent equation $x^2=0$ essentially), but this map is not an isomorphism. This means that the obstructions for the two problems
are in general not holomorphically equivalent and we must take care to distinguish between them.\footnote{
This structure plays a similarly important role for defining Donaldson-Thomas style invariants, where
the virtual class should essentially be the Euler class of the self-dual sub-bundle $E_+$ of the obstruction bundle.}

A disadvantage of the above formulation is that the expression $\kappa_+$ depends explicitly on the metric, but at the end of the day the superpotential can only depend on holomorphic data. A similar issue was addressed in the mathematics literature \cite{Oh:2020rnj}. The idea is that instead of the bundles $E_+$ and $E_-$, we should look at a holomorphic maximally isotropic sub-bundle $\Lambda$ and its dual $\Lambda^*$, which fit together in a sequence
\be
0 \ \to \Lambda \ \to E \ \to \Lambda^* \ \to \ 0
\ee
Then $\kappa$ can be written as a pair of sections $({\sf e},{\sf j})$ of $\Lambda$ and $\Lambda^*$ respectively, and the condition $Q(\kappa,\kappa) = 0$ can be written as ${\rm } {\sf e} \cdot {\sf j} = 0$. This formulation is familiar from two-dimensional field theories with $(0,2)$ supersymmetry \cite{Witten:1993yc}, which is not coincidental.\footnote{
 Indeed compactifying the heterotic string on a Calabi-Yau four-fold with a complex ASD bundle leads to a two-dimensional low energy effective theory with $(0,2)$ supersymmetry.}
Now imposing only ${\sf j} = 0$ does not imply that ${\sf e}$ and $\kappa$ vanish. However ${\sf e}$ defines a cosection $\Lambda^* \to \cO$, whose degeneracy locus is ${\sf e} = 0$, and the Euler class of $\Lambda^*$ (representing the zeros of ${\sf j}$) can be localized to ${\sf e} = 0$ \cite{Oh:2020rnj}. So at least in this sense the choice of metric is not relevant.

\newsubsection{Comparison with the open string $B$-model}

It's interesting to compare with the open string $B$-model \cite{Witten:1992fb}. Consider a form ${\sf A} = \sum_k {\sf A}^k$ where ${\sf A}^k \in \Omega^{0,k}({\rm End}_0({\cal W}))$ is a general $(0,k)$-form valued in ${\rm End}_0({\cal W})$. Then formally the open string topological $B$-model on a Calabi-Yau $5$-fold $Y$ seems to reduce to a field theory with action\footnote{
There are apparently some known issues with the cubic string field theory when the dimension differs from three which requires introducing additional
interactions (see \cite{Erler:2013xta}). We assume this will land on its feet as it would be hard to swallow if the string field theory for the open string $B$-model does not reproduce twisted versions of the supersymmetric Yang-Mills theories living on physical $D$-branes.}
\be\label{B_model_CS}
S \ = \ \int_Y \Omega^{5,0} \wedge {\rm Tr}({\sf A} \delb {\sf A} + {2\over 3} {\sf A} \wedge {\sf A} \wedge {\sf A})
\ee
Indeed this theory is known to be equivalent to the holomorphically twisted $10d$ supersymmetric Yang-Mills theory \cite{Baulieu:2010ch}].
As usual we can obtain the theory on lower dimensional branes by dimensional reduction, which replaces forms with anti-holomorphic indices
in the directions which were reduced to
sections of exterior powers of the normal bundle of the brane. When the normal bundle is trivial, this means that the twisted Yang-Mills theory on the brane can be expressed as a holomorphic Chern-Simons
theory where the fields live in \cite{Costello:2016mgj}
\be
{\sf A} \in \ \Omega^{(0,*)}(C^k)[\epsilon_1, \ldots, \epsilon_{5-k}] \otimes {\rm End}_0({\cal W})
\ee
and the $\epsilon_i$ denote Grassmann variables.
It is satisfying that the three natural candidates for a `topological' gauge theory on Calabi-Yau four-folds, namely the open string B-model holomorphic Chern-Simons theory,  the twisted $8d$ supersymmetric Yang-Mills theory and finally the complex version of topological Yang-Mills theory, are all apparently equivalent to each other.

\newpage

\newsection{Gauge theory on log Calabi-Yau spaces}
\seclabel{log_CY}

Let us now discuss how to adjust the gauge theory picture
for a log Calabi-Yau four-fold $(Y,Z)$. Recall that a log Calabi-Yau is a pair $(Y, Z)$, where $Y$ is a variety and $Z$ is an effective divisor in $Y$ such that the log canonical class $K_{(Y,Z)} \equiv K_Y +Z$ vanishes. 
Such pairs may be thought of as
non-compact Calabi-Yau spaces since by \cite{TianYau_I,TianYau_II} (and assuming smoothness) 
there exist a complete Ricci-flat K\"ahler metric on $Y$ that diverges along $Z$.
We refer to $Z$ as the log boundary. Note that it follows from the definition and the adjunction formula that $Z$ is itself a Calabi-Yau.

\newsubsection{Long exact sequence}

We start by expressing the relation
between the relevant bundle-valued Dolbeault
cohomology groups on $Y$ and those on $Z$.

We consider a holomorphic bundle ${\cal V}$ on $Z$ which may be extended to a holomorphic bundle ${\cal W}$ on $Y$.
In order to relate the cohomology groups on $Z$ with those of $Y$,
we use the short exact sequence on $Y$ given by
\be\label{res_sequence}
0 \ \to \ \cO(-Z) \ \to \cO \ \to \ \cO_Z \ \to \ 0
\ee
By tensoring with ${\rm End}_0({\cal W})$
we deduce the long exact sequence
\be\label{def_restriction_long}
\begin{array}{ccccccccc}
  0 & \to & H^0({\rm End}_0({\cal W})(-Z)) & {\stackrel{g_0}{\to}} & H^0({\rm End}_0({\cal W})) &  {\stackrel{h_0}{\to}}   & H^0({\rm End}_0({\cal V})) & & \\
   & {\stackrel{f_1}{\to}}  & H^1({\rm End}_0({\cal W})(-Z)) &  {\stackrel{g_1}{\to}} & H^1({\rm End}_0({\cal W})) & {\stackrel{h_1}{\to}}   & H^1({\rm End}_0({\cal V})) & & \\
   & {\stackrel{f_2}{\to}}   & H^2({\rm End}_0({\cal W})(-Z)) & {\stackrel{g_2}{\to}}  & H^2({\rm End}_0({\cal W})) & {\stackrel{h_2}{\to}}  & H^2({\rm End}_0({\cal V})) &  & \\
   & {\stackrel{f_3}{\to}}  & H^3({\rm End}_0({\cal W})(-Z)) &  {\stackrel{g_3}{\to}}   & H^3({\rm End}_0({\cal W})) &{\stackrel{h_3}{\to}}  & H^3({\rm End}_0({\cal V})) &  & \\
      & {\stackrel{f_4}{\to}}  & H^4({\rm End}_0({\cal W})(-Z)) &  {\stackrel{g_4}{\to}}   & H^4({\rm End}_0({\cal W})) & \to  & 0 &  &
\end{array}
\ee
This sequence is self-dual under Serre duality, which identifies $ H^{4-p}({\rm End}_0({\cal W}))$ with the dual of
$H^{p}({\rm End}_0({\cal W})(-Z))$.

The maps in (\ref{def_restriction_long}) are mostly obvious, but for later use let us discuss how to think about the coboundary maps $f_p$. Given an ${\rm End}_0({\cal V})$-valued $(0,p)$-form $\omega$ on $Z$ representing a class in $H^p({\rm End}_0({\cal V}))$, we can extend it to an ${\rm End}_0({\cal W})$-valued $(0,p)$-form $\widetilde{\omega}$ on $Y$, albeit one that is not necessarily annihilated by $\delb$. For example we could extend it to a form that vanishes on all but a small neighbourhood of the boundary $Z$. Then $\delb \widetilde{\omega}$ is an ${\rm End}_0({\cal W})$-valued  $(0,p+1)$-form on $Y$ that is obviously annihilated by $\delb$. It is of course zero as a class in $H^{p+1}({\rm End}_0({\cal W}))$, but it may represent non-zero class in $H^{p+1}({\rm End}_0({\cal W})(-Z))$ since $\delb\widetilde{\omega}$ vanishes along $Z$ but $\widetilde{\omega}$ does not vanish along $Z$. The map $f_p$ vanishes along classes that are obtained from restriction of $H^p({\rm End}_0({\cal W})) \to H^p({\rm End}_0({\cal V}))$, because such classes on $Z$ can obviously be extended to a class on $Y$ that is annihilated by $\delb$. 

\newsubsection{Hodge theory on tubular manifolds}
\subseclabel{LogCY_Hodge_Theory}

In order to proceed further we need to know what (if anything) the Dolbeault cohomology groups
in (\ref{def_restriction_long}) have to do with the fermionic
zero modes associated to the twisted gauge theory on our non-compact Calabi-Yau $Y$.
Recall from (\ref{Dirac_op}) that these are zero modes of a Dirac operator
$\delb^\dagger + \delb^+: \Lambda^{0,1} \to \Lambda^{0,0} \oplus \Lambda^{0,2}_+$ and its adjoint. As discussed we expect these zero modes to get some corrections because the $D$-terms we are interested in are not the standard ones, but this should not affect the asymptotics near the log boundary, as we should be able to recover the standard harmonic forms considered in the heterotic string near the boundary. Another reason to go through this in some detail is that one would hope that the $\star$-operator involved in the self-duality condition descends to Dolbeault cohomology. But since the natural non-degenerate quadratic form on a log Calabi-Yau pairs $H^2({\rm End}_0({\cal W}))$ with $H^2({\rm End}_0({\cal W}(-Z)))$,
this seems at first sight impossible as the $\star$-operator doesn't relate non-normalizable forms with `relative' forms that vanish at the log boundary. The resolution of this puzzle is rather interesting and fortunately does allow us to descend the $\star$-operator to Dolbeault cohomology on a log Calabi-Yau.

To understand the zero modes we need some results on Hodge theory on non-compact manifolds. The relevant results are currently only known for special classes of asymptotics, but that will give a decent sense of the issues. The asymptotic form of the Tian-Yau metric depends on the normal bundle $N_Y Z$ of $Z$ in $Y$. If the normal bundle is trivial (or torsion) then the metric is asymptotically cylindrical, i.e. we have
\be\label{tube_metric}
ds^2_Y\ \to \ dr^2 + d\theta^2 + ds^2_Z
\ee
and the corrections fall off exponentially as $e^{-\delta r}$ for some $\delta > 0$ \cite{Kovalev:2000,ACyl_CY:2012}. Here $(r, \theta)$ are coordinates on the cylinder ${\bf R}^+ \times S^1$. For the asymptotically cylindrical case essentially all the relevant results we need may be found in \cite{Melrose_APS}. This reference discussed real analogues of the problems at hand, but there are usually natural complex analogues and many of these were proven in \cite{ACyl_moduli:2014}.
See also \cite{Donaldson:2002}
for an overview of results on conventional four-dimensional instantons on asymptotically cylindrical manifolds.

If on the other hand the normal bundle $N_Y Z$ of $Z$ in $Y$ is positive, then the Tian-Yau metric takes a slightly more complicated asymptotic
form, for which explicit expressions may be found in \cite{Collins:2019}. Presumably one may prove analogous
Hodge theoretic results in this situation, but we regard the asymptotically cylindrical metrics as an interesting case study and restrict to them from now on.

It is convenient to use the complex coordinate
$z = e^{-r + i \theta}$. Then the metric and volume form on the cylinder take the form
\be
ds^2_{Cyl}\ =\ {dz d\bar{z}\over z \bar{z}}, \qquad dr \wedge d\theta \ = \ {i\over 2} {dz \wedge d\bar{z}\over z \bar{z}}
\ee
Similarly the logarithmic $(4,0)$-form is asymptotically of the form
\be
\Omega^{4,0}_Y \ = \ \Omega^{3,0}_Z \wedge {dz\over z}
\ee

As we saw in section \ref{Fermion_zero}, part of the equations for zero modes involves the vanishing of $\delb^+ \alpha$ for a $(0,1)$-form $\alpha$. We claim
that this is equivalent to requiring that $\delb\alpha = 0$. To see this first assume that $\alpha$ is integrable
and consider the expression
\be
\int_Y \Omega^{4,0} \wedge {\rm Tr}(\delb\alpha \wedge \delb\alpha)\ \propto\ \int_Z \Omega^{3,0} \wedge {\rm Tr}(\alpha\wedge \delb \alpha)\ =\ 0
\ee
On the other hand decomposing into holomorphic self-dual and anti-self-dual parts we have
\be
\int_Y \Omega^{4,0} \wedge {\rm Tr}(\delb\alpha \wedge \delb\alpha)\  =\ || \delb^+\alpha||^2 - ||\delb^-\alpha||^2
\ee
where $|| \beta||^2 = \int_Y \Omega^{4,0} \wedge {\rm Tr}(\beta \wedge \star \beta)$, and the claim follows. We will also
be interested in zero modes which are not integrable, but which are still bounded as $r \to \infty$ and such that $\delb\alpha$ and $\delb^\dagger\alpha$ are integrable, so the above argument still applies.
Thus we can work with $\delb$ instead of $\delb^+$, and effectively we will be interested in forms that are simultaneously annihilated by $\delb$ and $\delb^\dagger$.

On a non-compact space we have to distinguish between the harmonic forms, which satisfy $\Delta_{\delb}\alpha = 0$, and the forms which satisfy $\delb\alpha = \delb^\dagger\alpha  = 0$. The latter are sometimes referred to as the harmonic fields, and we will borrow this terminology. At first sight the fermionic zero modes look to be related to harmonic fields, but it turns out the harmonic forms also have a role to play.

Due to the simplicity of (\ref{tube_metric}), the large $r$ behaviour of harmonic forms is not hard to understand. Using separation of variables, harmonic forms
for  (\ref{tube_metric}) are superpositions of functions of the form
\be
R_k(r)\, \Theta_k(\theta, w)
\ee
where $w$ denotes coordinates on $Z$, $\Theta(\theta, w)$ is an eigenfunction of the Laplacian on $S^1 \times Z$ with eigenvalue $-k^2 \leq 0$, and $R(r)$ is an eigenfunction of $\del^2 /\del r^2$. Normalizable solutions fall off like $e^{-kr}$ or faster, where $k > 0 $ and $-k^2$ is the smallest non-zero eigenvalue of the Laplacian on  $S^1 \times Z$.

We can ask for the $L^2$-normalizable harmonic fields, which we denote as ${\bf H}_{L^2}^k(Y,{\rm End}_0({\cal W}))$.
As these forms are integrable we may apply the usual argument to show that harmonic fields are harmonic forms and vice versa. It may be shown that in terms of Dolbeault cohomology this space may be identified with the image of the map
$g_k$ in (\ref{def_restriction_long}). In order to get Hodge theoretic representatives for the remaining Dolbeault generators we also have to consider non-normalizable forms. Let us first define the space of `extended' harmonic forms as
\be
{\rm null}_-^k(\Delta_{\delb}) \ = \ \{\omega \in \Omega^{0,k}({\rm End}_0({\cal W}))\, |\, \Delta_{\delb}\omega = 0\ {\rm and}\ (z\bar{z})^\epsilon\omega\ {\rm integrable \ for \ small}\ \epsilon > 0 \}
\ee
The point of this is that we include non-normalizable harmonic forms which behave asymptotically as $R_0(r) \sim a + br$ for constants $a$ and $b$, but still exclude forms which behave as 
$e^{kr}$ for $k > 0$. 
Of these, the forms which behave as $R_0(r) \sim a$ (i.e. $b=0$) are exactly the ones that are also annihilated by $\delb + \delb^\dagger$, which is what
we are interested in. We may further
subdivide them according to their asymptotic behaviour as
\ba
{\bf H}_{rel}^q({\rm End}_0({\cal W})) \is \{ \omega^q \in {\rm null}_-^{q}(\Delta_{\delb})\, |\, \omega^q \sim \alpha^{q-1} \wedge {d \bar{z}\over \bar{z}} + {\rm normalizable} \}\\
{\bf H}_{abs}^q({\rm End}_0({\cal W})) \is \{ \omega^q \in {\rm null}_-^{q}(\Delta_{\delb})\, |\, \omega^q \sim \alpha^{q} + {\rm normalizable} \}
\ea
where $\alpha^q \in {\bf H}^q_{L^2}(Z,{\rm End}_0({\cal V}))\cong H^q(Z,{\rm End}_0({\cal V}))$.
We refer to these as the relative and the absolute Hodge groups. Note the asymptotics are
reminiscent of Dirichlet and Neumann boundary conditions on a manifold where the boundary is at finite distance.
As the notation suggest, these spaces of harmonic fields turn out to be isomorphic
to the relative and absolute Dolbeault cohomology groups:
\ba
{\bf H}_{rel}^q({\rm End}_0({\cal W})) & \cong & H^q(Y,{\rm End}_0({\cal W})(-Z))  \\
{\bf H}_{abs}^q({\rm End}_0({\cal W})) & \cong & H^q(Y,{\rm End}_0({\cal W}))
\ea
Furthermore we expect the Hodge groups to admit the following orthogonal decomposition:\footnote{
The second of these is shown in \cite{ACyl_moduli:2014}; the first, although the natural complex analogue
of proposition (6.16) in \cite{Melrose_APS}, does not seem to have been proven.}
\ba
{\bf H}_{rel}^q({\rm End}_0({\cal W})) \is {\bf H}_{L^2}^q({\rm End}_0({\cal W})) \oplus \delb\, {\rm null}_-^{q-1}(\Delta_{\delb}) \\
{\bf H}_{abs}^q({\rm End}_0({\cal W})) \is {\bf H}_{L^2}^q({\rm End}_0({\cal W})) \oplus \delb^\dagger {\rm null}_-^{q+1}(\Delta_{\delb})
\ea
This decomposition allows us to write the Hodge theoretic implementations of the maps in our long exact sequence (\ref{def_restriction_long}) of Dolbeault cohomology groups. We already claimed that the space the $L^2$-normalizable harmonic forms is isomorphic to the image
of the maps $g_k$. Using the possible asymptotic behaviours of harmonic forms $\omega^{q+1} \in {\rm null}_-^{q+1}(\Delta_{\delb})$, we see that
$\delb^\dagger \omega^{q+1}$ is of the form
\be\label{Abs_Dolbeault_asymptotic}
\delb^\dagger \omega^{q+1}\ =\ \alpha^{q} + \delb^\dagger( {\rm normalizable})
\ee
for $\alpha^q \in {\bf H}^q_{L^2}({\rm End}_0({\cal V}))$. Such forms $\delb^\dagger \omega^{q+1}$ are annihilated by $\delb$, $\delb^\dagger$ and completely determined by $\alpha^{q}$.
Thus this allows us to define a map $h_k: {\bf H}^k_{abs}(Y,{\rm End}_0({\cal W})) \to {\bf H}^k_{L^2}(Z,{\rm End}_0({\cal V}))$, by mapping
to the asymptotically limiting form $\alpha^{k}$. Similarly we have
\be\label{Rel_Dolbeault_asymptotic}
\delb\omega^{q-1}\ =\ \alpha^{q-1} \wedge {d\bar{z}\over \bar{z}} + \delb ({\rm normalizable})
\ee
for $\alpha^{q-1} \in {\bf H}^{q-1}({\rm End}_0({\cal V}))$, completely determined by $\alpha^{q-1}$. Furthermore as in \cite{Melrose_APS} we expect that we can split
$ {\bf H}^k_{L^2}({\rm End}_0({\cal V}))$ into two orthogonal subspaces $H^k_1 \oplus H^k_2$, such that forms in $H^k_1$ can be extended to harmonic forms
on $Y$ with constant asymptotics as in (\ref{Abs_Dolbeault_asymptotic}), and forms in $H^k_2$ can be extended to harmonic forms on $Y$ with logarithmic growth, so that after applying $\delb$ we get a harmonic form with asymptotics of the form (\ref{Rel_Dolbeault_asymptotic}). Then $H^k_1$ is the image of $h_k$, and we can define a map $f_{k+1}: {\bf H}^{k}_{L^2}(Z,{\rm End}_0({\cal V})) \to {\bf H}^{k+1}_{rel}(Y,{\rm End}_0({\cal W}))$ which vanishes on $H^k_1$. This recovers all the maps in our long exact sequence (\ref{def_restriction_long}).

Let us consider the action of the $\star$-operator. The two types of non-normalizable
forms are exchanged by $\star$, which acts on the asymptotics as
\be
\alpha^{q-1} \wedge {d\bar{z}\over\bar{z} }\ \longleftrightarrow\ \star_3\, \alpha^{q-1}
\ee
with $\star_3 = \bar{*}_3(\Omega^{3,0} \wedge \cdot)$ the corresponding operator on the Calabi-Yau three-fold $Z$. The $\star$-operator further
maps normalizable harmonic forms to normalizable harmonic forms. 

Note the curiosity that part of the relative cohomology (namely the image of $f_k$) gets associated to non-normalizable harmonic fields, despite the fact that the relative cohomology is constructed from forms that vanish at the boundary. Nevertheless, as in \cite{Melrose_APS}, there are natural maps that give the expected isomorphisms. Let us first review the coboundary map for the Dolbeault cohomology groups. Given a class $\alpha^{q-1}$ in $H^{q-1}(Z, {\rm End}_0({\cal V}))$, we can always extend it to a $(0,q-1)$-form $\tilde\alpha^{q-1}$ on $Y$. For example we may choose a bump function $\phi(r)$ which is supported only close to $z=0$ and equal to $1$ in a small neighbourhood of $z= 0$, and define $\tilde\alpha^{q-1} = \phi\cdot \alpha^{q-1}$. Then $\delb\tilde\alpha^{q-1} = \delb\phi \cdot \alpha^{q-1}$ is class in the relative cohomology $H^{q}(Y, {\rm End}_0({\cal W})(-Z))$, as we clearly have
$\delb^2\tilde\alpha^{q-1}=0$ and $\delb\tilde\alpha^{q-1}|_Z = 0$. If we choose another extension over $Y$ then the difference between
the corresponding $\tilde \alpha^{q-1}$'s vanishes along $Z$ and hence we get the same Dolbeault cohomology class. The kernel of the coboundary map consists of classes $\alpha^{q-1}$ on $Z$ which admit an extension to a closed $(0,q-1)$-form on $Y$.

Now recall that the coboundary map for the Hodge cohomology takes a harmonic form $\alpha^{q-1}$ on $Z$ and lifts it to a harmonic field on $Y$ of the form (\ref{Rel_Dolbeault_asymptotic}). For a suitable bump function $\psi$
we may consider the map
\be\label{rel_Hodge_Dolbeault_map}
\delb\omega^{q-1}\ \to\ \tilde{\omega}^q\ =\ \delb\omega^{q-1} - \delb (\psi(z\bar{z}) \log(z\bar{ z}) \alpha^{q-1})
\ee
If we choose $\psi$ appropriately then we remove the non-normalizable piece and reproduce the coboundary map for Dolbeault cohomology (plus an extra exact piece which vanishes along $Z$, which doesn't affect the cohomology class). This relates the relevant pieces in $H^q(Y,{\rm End}_0({\cal W})(-Z))$ and ${\bf H}^q_{\rm rel}(Y,{\rm End}_0({\cal W}))$. The choice of the function $\psi$ is irrelevant as long as we fix the behaviour as $z\to 0$.

For the purpose of computing correlation functions we are interested in the non-normalizable modes, as the normalizable ones would be gauge exact in the DFM model and therefore not physical.
But in fact it appears that when ${\cal W}$ is a tautological bundle on a geometry of the type considered in section \ref{Compactified_correspondence}, then the maps $g_k$ are typically zero and hence in particular there would be no normalizable modes anyways. The reason for this is that the bundle ${\cal W}$ is constructed as a Fourier-Mukai transform of a spectral sheaf, and
hence we expect Parseval identities of the form
\be
\Ext^k_Y({\cal W},{\cal W}) \ \cong \ \Ext^k_C(L, L)
\ee
where $L$ is the spectral sheaf supported on the spectral cover $C$. But $\Ext^k_C(L, L)$ maps injectively to
$\Ext^k_Z(i_{C*}L, i_{C*}L) \cong \Ext^k_Z({\cal V},{\cal V})$. Restricting to the traceless parts, we see that
when these Parseval identities hold the maps
$h_k$ are injective and hence the maps $g_k$ vanish.
In the case where the spectral cover is smooth and irreducible and $L$ is a line bundle
we show explicitly in \cite{MF_Extension} that the Parseval identity holds for $k=1$ and hence the map $g_1$ is zero. By a similar calculation we can probably also show that $g_2$ vanishes, and the map $g_0$ vanishes trivially as ${\rm End}_0{\cal W}$ has no holomorphic sections. By Serre duality the remaining maps $g_i$ would then also vanish.

\newsubsection{Combined deformation of bundle and complex structure}
\subseclabel{Bulk_boundary_compare}

Recall that in the heterotic string,  $H^0({\rm End}_0({\cal V}))$ counts unbroken gauge symmetries and $H^1({\rm End}_0({\cal V}))$ counts
charged matter fields (plus bundle moduli). From the long exact sequence we see that these get related to bundle-valued Dolbeault cohomology groups on $Y$, and because the sequence is exact we recover all the Dolbeault cohomologies of ${\cal V}$ this way, in particular we recover the degrees of freedom that are usually attributed to $M2$-branes wrapped on vanishing cycles. We also saw that the Dolbeault cohomologies of ${\cal V}$ should correspond precisely to the non-normalizable Dolbeault cohomologies on $Y$; this matches with the expectation from the DFM model that only defect changing or boundary changing modes can yield additional degrees of freedom beyond the usual degrees of freedom of the three-form. 

On the other hand, from a more purely $M$-theory or $F$-theory perspective, some of these Dolbeault cohomology groups on $Y$ don't seem all that natural. A more natural way to state the problem is that we have a complex manifold $Y$ with a holomorphic bundle ${\cal W}$ on top of it, and we are interested in the joint deformation problem for the pair $(Y, {\cal W})$. Fortunately this is also a well-known deformation theory problem, see \cite{Huang:1995} and also the review \cite{ChanSuen:2014} and references therein. This gives an alternative view of the degrees of freedom which will be useful when we look at the superpotential. We will briefly review some aspects and then compare with the cohomology groups obtained from the long exact sequence for the log Calabi-Yau models  of section \ref{Compactified_correspondence}.

Deformations  of the pair $(Y, {\cal W})$ are controlled by the Atiyah bundle, which is defined as an extension
\be
 0 \to {\rm End}_0({\cal W}) \to {\cal A} \to TY \to 0
\ee
As a $C^\infty$ bundle, ${\cal A}$ is just the direct sum $ {\rm End}_0({\cal W}) \oplus TY$, so the set of $(0,p)$-forms $\omega^{0,p} \in \Omega^{0,p}({\cal A})$ consists of pairs $(a^{0,p}, \xi^{0,p})$ where $a^{0,p} \in \Omega^{0,p}({\rm End}_0({\cal W}))$ and
$\xi^{0,p} \in \Omega^{0,p}(TY)$. 
However as a holomorphic bundle ${\cal A}$ is a non-trivial extension defined by the Atiyah class, i.e the curvature ${\sf F}^{1,1}$ interpreted as a class in $H^1(TY^* \otimes {\rm End}_0({\cal W}))$. Concretely this means that we use the deformed Dolbeault operator
\be
\delb_{\sf F} \ = 
\left(
\begin{array}{cc}
\delb & {\sf F} \\
0 & \delb \\
\end{array}
\right)
\ee
More explicitly, we have
\be
\delb_{\sf F}(a,\xi) \ = \ (\delb a + D_\xi {\sf F},\delb \xi)
\ee
where $(D_\xi {\sf F} )_{\bar i_1 \ldots \bar i_{p+1}}= \xi_{\bar i_1 \dots \bar i_p}^j {\sf F}_{j \bar i_{p+1}}$. This Dolbeault operator satisfies $\delb_{\sf F}^2 = 0$, so we can use it to define the cohomology of $(0,p)$-forms valued in ${\cal A}$, denoted as usual by
$H^p(Y, {\cal A})$. There is a bracket operation given by
\be\label{DLGA_bracket}
[ (a, \xi), (a', \xi' )]  \ = \ ( [a, a'] + D_{\xi'} \del_A a - D_{\xi} \del_A a' ,[\xi, \xi'])
\ee
which satisfies the Jacobi identity. For $p$-forms $\omega^{0,p} = (a, \xi)$ we have
\be
\delb_{\sf F} \,[\omega^{0,p},\omega'^{0,q}] = [\delb_{\sf F}\, \omega^{0,p},\omega'^{0,q}] + (-1)^p [\omega^{0,p}, \delb_{\sf F}\,  \omega'^{0,q}]
\ee
The triple $(\Omega^{0,*}({\cal A}), \delb_{\sf F}, [,])$ defines a differential graded Lie algebra, usually considered the natural setting for
deformation theory problems. 

Combined deformations of bundle and complex structure are
governed by the Maurer-Cartan equation:
\be
\delb_F \omega^{0,1} + \half [\omega^{0,1}, \omega^{0,1}] \ = \ 0
\ee
The first order deformations are classified by $H^1(Y, {\cal A})$, and the obstructions will live in  $H^2(Y, {\cal A})$.

Let us unpack why the Dolbeault cohomology group $H^1(Y, {\cal A})$ classifies first order deformations of the pair $(Y, {\cal W})$.  Under a complex structure deformation, the curvature of the
bundle ${\cal W}$ may develop a $(0,2)$-part, and so might no longer be holomorphic. Under a perturbation ${\sf A} \to {\sf A} + \epsilon\delta {\sf A} + \epsilon^2 \delta {\sf B} + \cO(\epsilon^3)$ the curvature changes as
${\sf F} \to {\sf F} + \epsilon d_{\sf A}\delta {\sf A} + \epsilon^2(d_A \delta {\sf B} + [\delta{\sf A},\delta{\sf A}]) + \cO(\epsilon^3)$. Under a combined perturbation of both
the gauge field and the complex structure, to first order the $(0,2)$-part transforms as
\be
{\sf F}^{0,2} \ \to {\sf F}^{0,2} + \delb \delta{\sf A}^{0,1} + D_\xi {\sf F}^{1,1} \ = \  {\sf F}^{0,2}  + \delb_{\sf F}(\delta{\sf A}^{0,1}, \xi^{0,1})
\ee
where $(D_\xi {\sf F})_{\bar i \bar k} =  \xi_{\bar i}^j{\sf F}_{j\bar k}$ and $\xi_{\bar i}^j \in H^1(TY)$. For the bundle to remain holomorphic after a complex structure deformation, we need to keep the $(0,2)$ part vanishing, and hence we see that $(\delta{\sf A}^{0,1}, \xi^{0,1})$ should be annihilated by $ \delb_{\sf F}$. Modding out by the image of $ \delb_{\sf F}$ just corresponds to modding out by infinitesimal gauge and complex coordinate transformations. 

Now when $Y$ is a log Calabi-Yau we are not interested in all complex structure deformations of $Y$, but in deformations which preserve $Z$, i.e. deformations of the log pair $(Y,Z)$, so in this case we will replace $TY$ by the logarithmic tangent bundle $TY(-\log Z)$. It fits in an exact sequence
\be
0 \to TY(-Z) \ \to TY(-\log Z) \ \to \ T_Z \ \to \  0
\ee
Infinitesimal complex structure deformations of the pair $(Y,Z)$ are classified by the cohomology group $H^1(TY(-\log Z))$, and there are Tian-Todorov type theorems which say that these deformations are all unobstructed \cite{Iacono:2013,Katzarkov:2014bma}. From the associated long exact sequence, we see that these are built up from deformations that leave $Z$ completely fixed (coming from $H^1(TY(-Z))$), and complex structure deformations of $Z$ itself (coming from $H^1(TZ)$).  In the constructions of section. \ref{Compactified_correspondence} , since $Y$ is obtained as a blow-up of the spectral cover, the deformations in from $H^1(TY(-Z))$ come precisely from deformations of the spectral cover $C \subset Z$. This motivates focusing
on the complex structure deformations in $H^1(TY(-Z))$ that leave $Z$ completely fixed. 
Accordingly we consider a holomorphic sub-bundle ${\cal A}_Z \subset {\cal A}$ of the Atiyah bundle, given by
\be\label{Atiyah_log_CY}
 0 \to {\rm End}_0({\cal W}) \to {\cal A}_Z \to TY(-Z) \to 0
\ee

Let us now consider the joint deformations of $(Y,{\cal W})$ keeping $Z$ fixed, counted by $H^1({\cal A}_Z)$. 
From the associated long exact sequence we get
\be
 0 \to H^1({\rm End}_0({\cal W}) )\to H^1({\cal A}_Z) \to H^1(TY(-Z))  \to H^2({\rm End}_0({\cal W}) )
\ee
where we assumed that $H^0(TY(-Z))=0$, as expected for a log Calabi-Yau. As we saw the image of $H^1({\rm End}({\cal W})(-Z) )\to H^1({\rm End}({\cal W}))$ (if non-zero) should consist of normalizable modes that are exact in the DFM model, so we mod out by these to get a reduced cohomology that we denote by $\widetilde H^1(Y, {\cal A}_Z)$. 

Consider an ${\cal A}_Z$-valued $(0,1)$-form $(\delta{\sf A}, \xi)$ representing a class in $\widetilde H^1(Y, {\cal A}_Z)$. Restricting to $Z$, we get a combined complex structure/bundle deformation on $Z$. However since $\delb \xi = 0$ and $\xi$ actually represents a class in $H^1(TY(-Z))$, it vanishes along $Z$, hence we also have $\delb \delta{\sf A} = 0$ and we get a map to $\widetilde H^1(Y, {\cal A}_Z) \to H^1({\rm End}_0({\cal V}) )$. It is in general not clear that this map is injective. Injectivity would fail if there exist a pair $(\delta{\sf A}, \xi)$ such that $\xi$ is a non-zero class in $H^1(TY(-Z))$ but $\delta{\sf A}|_Z$ is zero in $H^1({\rm End}_0({\cal V}) )$. However for the geometries considered in section 
 \ref{Compactified_correspondence}, by construction any complex structure deformation in $H^1(TY(-Z))$ comes from a deformation of the equation of the spectral cover, hence (by Fourier-Mukai) from a bundle deformation in $H^1({\rm End}_0({\cal V}) )$. Hence in our situation it seems reasonable to assume that the map $\widetilde H^1(Y, {\cal A}_Z) \to H^1({\rm End}_0({\cal V}) )$ is an injection. 

It's easy to see that in general we cannot expect to recover all of $H^1({\rm End}_0({\cal V}) )$ from $\widetilde H^1(Y, {\cal A}_Z)$. Essentially, on the $F$-theory side we've introduced an asymmetry by the choice of resolution of $Y$, whereas the pair $(Z, {\cal V})$ does not see this asymmetry. (This asymmetry does play a role in the $M$-theory version, as we briefly discussed in section ..).  However by applying Serre duality we can still relate Dolbeault cohomology of ${\cal V}$  to some Dolbeault cohomology of ${\cal A}_Z$. Let us illustrate this for simplicity when $Z$ is just an elliptic curve $E$ and 
${\cal V} = \cO_E^{\oplus n}$. The corresponding rational surface has an $A_{n-1}$ singularity, and we lift the bundle (up to a twist) as 
\be
{\cal W} \ = \ \cO_S(\alpha_1 + \ldots + \alpha_{n-1}) \oplus \ldots \oplus \cO_S(\alpha_1) \oplus \cO_S
\ee
where the $\alpha_i$ are effective curves.
The upper triangular generators of $H^1({\rm End}_0({\cal V}))$ are inherited from $H^1({\rm End}_0({\cal W}))$. The lower triangular generators of $H^1({\rm End}_0({\cal V}))$ map to classes in $H^2({\rm End}_0({\cal W}(-Z)) $ that are not related to complex structure deformations, but their Serre duals comprise the upper triangular generators of the inclusion 
$H^{0}({\rm End}_0({\cal W})) \to 
H^{0}({\rm End}_0({\cal V}))$, which will hence show up in $H^0(Y, {\cal A}_Z)$.
Finally the diagonal generators of $H^1({\rm End}_0({\cal V}))$
map to bundle/complex structure moduli that deform the singularity; their Serre duals can also be understood as the diagonal generators in the inclusion $H^{0}({\rm End}_0({\cal W})) \to 
H^{0}({\rm End}_0({\cal V}))$.

\newsubsection{Chirality}

Finally we note that we are often interested in a situation where ${\cal V}$ or ${\cal W}$ is reducible. Then we get chiral representations of the unbroken gauge group from forms valued in $\Hom({\cal W}_1, {\cal W}_2)$ and we can ask
for the net amount of chiral matter. By tensoring (\ref{res_sequence}) with $\Hom({\cal W}_1, {\cal W}_2)$ we get a long exact sequence analogous to (\ref{def_restriction_long}). One simple consequence of that sequence is that
\be
\chi({\cal V}_1,{\cal V}_2) \ = \ \chi({\cal W}_1, {\cal W}_2) - \chi({\cal W}_1, {\cal W}_2(-Z))
\ee
Further note that by Serre duality we have
\be
\chi({\cal W}_1, {\cal W}_2(-Z)) \ = \ \chi({\cal W}_2, {\cal W}_1)
\ee
But expressions of the form $\chi({\cal V}_1,{\cal V}_2)$  on $Z$ exactly count the net amount of chiral matter associated to chiral representations of the unbroken gauge group. By the above equality and an application of the index theorem on $Y$ we may therefore write this quantity as an expression in terms of characteristic classes of bundles and Todd classes on $Y$.
On the other hand, we already know that we can compute the net amount of chiral matter in terms of integrals of the ${\sf G}$-flux on $Y$ \cite{Donagi:2008ca}. We hope to explore this further \cite{F_Fluxes}.

\newpage

\newsection{Holomorphic couplings as obstructions}
\seclabel{Obstructions}

One of the original motivations for this project was the lack of a proper and intrinsic definition of Yukawa couplings and higher order holomorphic couplings
for charged matter in $M$/$F$-theory as some kind of obstruction. Since we now have recovered a more-or-less standard deformation theoretic framework, we can see how
such a description arises. 

We have already reviewed in section  \ref{Local_moduli} how the
deformation theory of sheaves and bundles on a fixed Calabi-Yau four-fold is governed by a holomorphic potential function $\Phi_{BV}$. This is defined on fields and anti-fields, so needs to be restricted to
physical fields in order to get the superpotential. As reviewed in appendix \ref{Deformation_Theory}, if we pick a Kuranishi map $\kappa$ such  that $Q(\kappa, \kappa) = 0$, then the couplings are obtained in the standard way by expanding the Kuranishi map as a sum of Massey products $ {\sf m}_k$:
\be
W(\delta A^{0,1}, B^{0,2}_+) \  =\  \sum_{k \geq 2}\  \int_{Y_4} \Omega^{4,0} \wedge {\rm Tr}( {\sf m}_k(\delta A^{0,1}, \ldots ,\delta A^{0,1}) \wedge B^{0,2}_+)
\ee
where $\delta A^{0,1} \in H^1({\rm End}_0({\cal W}))$ and $B^{0,2}_+ \in H^2_+({\rm End}_0({\cal W}))$. The equations of motion recover the condition that $\Pi_+ \circ \kappa = 0$, which is the statement that the deformed connection satisfies the complex ASD equation
${\sf F}^{0,2}_+ = 0$.  

In our setting we would like to make two adjustments. As we saw in section \ref{Bulk_boundary_compare}, it's more natural to consider the combined deformation problem for the pair $(Y, {\cal W})$. The first order deformations of the pair $(Y, {\cal W})$ are classified by $H^1(Y, {\cal A})$, and we 
can construct a Kuranishi map \cite{ChanSuen:2014}
\be
\kappa : U \subset H^1(Y, {\cal A}) \ \to \ H^2(Y, {\cal A})
\ee
which measures the obstruction to a finite solution to the Maurer-Cartan equation, i.e. a simultaneous deformation of the pair $(Y, {\cal W})$. Here $U$ is a small open subset containing the origin. This generalized version can also be captured with a superpotential.

The second adjustment is that we want to mod out by the remnant one-form gauge transformations, i.e. we want to further mod out by deformations $\delta A^{0,1}$ that are exact when restricted to the exceptional locus. Unfortunately, even though the original superpotential $\int_Y \Omega \wedge {\rm Tr}({\sf F} \wedge {\sf F})$ is clearly invariant under one-form gauge transformations, we still have some technical issues writing a general effective superpotential that is invariant. To simplify, we will assume for now that we keep the complex structure of $Y$ fixed, and we only consider the quadratic term in $\kappa$, i.e. we consider
\be
W(\delta A^{0,1}, B^{0,2}_+)_{\rm Yukawa} \  =\   \int_{Y_4} \Omega^{4,0} \wedge {\rm Tr}( [\delta A^{0,1},\delta A^{0,1}] \wedge B^{0,2}_+)
\ee
It's easy to see that if $a \in H^1({\rm End}_0({\cal W}))$ and we add a class 
$\delta a \in H^1({\rm End}_0({\cal W}))$ that vanishes along the exceptional locus, then $[a + \delta a, a + \delta a] = [a,a]$ plus terms that vanishes when restricted to the exceptional locus. So we can make the superpotential invariant under such one-form gauge transformations by requiring $B$ to integrate to zero against classes in $H^2({\rm End}_0({\cal W}))$ that restrict to zero along the exceptional locus.

For a log Calabi-Yau $(Y,Z)$  we need to make some small adjustments to the above formula. The Yoneda product of two generators in $H^1({\rm End}_0({\cal W}))$ lives in $H^2({\rm End}_0({\cal W}))$ and consequently by Serre duality $B$ will live in $H^2({\rm End}_0({\cal W})(-Z))$. One-form gauge transformations in $H^1({\rm End}_0({\cal W}))$ in principle should vanish both along the boundary $Z$ and along the exceptional locus. However as we discussed previously, for the geometries encountered in heterotic/$F$-theory duality, it appears to be sufficient to impose vanishing 
along $Z$. Then to make the superpotential gauge invariant, we require $B$ to pair to zero with the image of $H^2({\rm End}_0({\cal W})(-Z)) \to H^2({\rm End}_0({\cal W}))$. But this is just another way of saying that $B$ is in the image of the coboundary map $H^1({\rm End}_0({\cal V})) \to H^2({\rm End}_0({\cal W})(-Z))$. To prove this, first note that due to the long exact sequence and due to the fact that $\dim H^2({\rm End}_0({\cal W})(-Z)) = \dim  H^2({\rm End}_0({\cal W}))$, the two spaces have the same dimension, so if we can show that one space is contained in the other then the result follows. Now if $B$ is in the image of the coboundary map then we can write
$B = \delb a^{0,1}$ where $a^{0,1}|_Z$ is a class in $H^1({\rm End}_0({\cal V}))$. Integrating against an element $g^{0,2} \in H^2({\rm End}_0({\cal W}))$ gives
\be
\int_Y \Omega^{4,0} \wedge {\rm Tr}(g^{0,2} \wedge \delb a^{0,1}) \ = \ -2\pi i \int_Z \Omega^{3,0} \wedge {\rm Tr}(g^{0,2} \wedge a^{0,1}) 
\ee
where we used $\delb (1/z) = 2\pi i \delta(z,\bar{z})$.
Thus if $g^{0,2}$ is in the image of $H^2({\rm End}_0({\cal W})(-Z)) \to H^2({\rm End}_0({\cal W}))$ then the integral vanishes, so $B$ does indeed integrate to zero against classes of this form. We conclude that for gauge invariance we should take $B$ to be in the image of the coboundary map. 

The $\star$-operator takes $H^2({\rm End}_0({\cal W})(-Z))$ to $H^2({\rm End}_0({\cal W}))$, so to make $B$ self-dual we should also add to it a piece in $H^2({\rm End}_0({\cal W}))$, again integrating to zero against the image of $H^2({\rm End}_0({\cal W})(-Z)) \to H^2({\rm End}_0({\cal W}))$. But such a piece automatically integrates to zero against classes in 
$H^2({\rm End}_0({\cal W}))/ im(H^2({\rm End}_0({\cal W})(-Z)) \to H^2({\rm End}_0({\cal W})))$ (as the trace of the product ends in $H^4(\cO)$, which is zero on a log Calabi-Yau), so we don't need to add it explicitly.

It follows that we may write the Yukawa couplings on the log four-fold $(Y,Z)$ as 
\be
W_{\rm Yuk} \ = \ \int_Y \Omega^{4,0} \wedge {\rm Tr}([a,a] \wedge \delb a^{0,1})
\ee
Integrating by parts yields
\be
W_{\rm Yuk} \ =\ -2 \pi i \int_Z \Omega^{3,0} \wedge {\rm Tr}([a,a] \wedge a^{0,1})
\ee
which agrees with the standard heterotic Yukawa coupling (\ref{CS_Yukawa}).

\noindent{\it Acknowledgements:}
We are grateful to D.~Freed, T.~Pantev and E.~Witten for discussion, and Y.~Cao and N.~Leung for correspondence.
Special thanks to Francis and Julian. The research of RD was partially supported by NSF grant DMS 2001673; 
by NSF grant DMS 2244978, FRG: Collaborative Research: New birational invariants; 
and by Simons Foundation Collaboration grant \#390287 “Homological Mirror Symmetry”.

\newpage

\appendix

\renewcommand{\newsection}[1]{
\addtocounter{section}{1} \setcounter{equation}{0}
\setcounter{subsection}{0} \addcontentsline{toc}{section}{\protect
\numberline{\Alph{section}}{{\rm #1}}} \vglue .6cm \pagebreak[3]
\noindent{\bf  \thesection. #1}\nopagebreak[4]\par\vskip .3cm}
\renewcommand{\newsubsection}[1]{
\addtocounter{subsection}{1}
\addcontentsline{toc}{subsection}{\protect
\numberline{\Alph{section}.\arabic{subsection}}{#1}} \vglue .4cm
\pagebreak[3] \noindent{\it \thesubsection.
#1}\nopagebreak[4]\par\vskip .3cm}

\newsection{Appendix: Elements of deformation theory}
\seclabel{Deformation_Theory}

We describe some aspects of the deformation theory of holomorphic bundles. A bundle with connection ${\sf A}$ is
holomorphic precisely when ${\sf F}^{0,2}$ vanishes. By choosing the Chern connection we may assume that ${\sf A}^{0,1} = 0$.
Now suppose we try to deform the connection to ${\sf A} + {\sf a}$. Then we find that ${\sf a}$ has to satisfy the Maurer-Cartan equation
\be\label{hol_MC}
{\sf F}^{0,2}_{A+a} \ = \  \delb {\sf a} + [{\sf a}, {\sf a}] \equiv 0
\ee
To fix the gauge invariance $ {\sf a} \to {\sf a} + \delb \lambda$ we impose $\delb^\dagger {\sf a} = 0$. The Hodge decomposition
is the orthogonal decomposition given by
\be
\Lambda^{0,p}({\rm End}_0({\cal W})) \ = \ H^{0,p}({\rm End}_0({\cal W}))  \oplus \delb \Lambda^{0,p-1}({\rm End}_0({\cal W}))  \oplus \delb^\dagger \Lambda^{0,p+1}({\rm End}_0({\cal W}))
\ee
Let ${\cal H}, \Pi_{\delb}$ and $ \Pi_{{\delb}^\dagger}$ denote the projections of these three pieces.
From (\ref{hol_MC}) it follows that $\delb [{\sf a},{\sf a}] = 0$ and hence that
$[{\sf a},{\sf a}]\in H^{0,2} \oplus {\rm Im}( \delb)$. Writing
\be
[{\sf a},{\sf a}] = {\cal H}([{\sf a},{\sf a}]) + \Pi_{\delb}([{\sf a},{\sf a}])
\ee
it is clear we can solve (\ref{hol_MC}) precisely when ${\cal H}([{\sf a},{\sf a}]) = 0$. Furthermore
${\sf a}$ will satisfy the equation
\be\label{simplified_MC}
\delb {\sf a} + \Pi_{\delb}([{\sf a},{\sf a}]) = 0
\ee

In order to solve the above equation we consider a power series expansion
\be
{\sf a}\ =\ {\sf a}_1 + {\sf a}_2 + ...
\ee
To first order the equations $\delb {\sf a}_1 = {\delb}^\dagger {\sf a}_1 = 0$ imply that
${\sf a}$ is a harmonic representative of
$H^{0,1}$. To proceed, note that $\delb$ is an isomorphism $\delb^\dagger \Lambda^{0,p+1} \to
\delb \Lambda^{0,p}$, so there exist an inverse $1/ \delb : \delb \Lambda^{0,p} \to \delb^\dagger \Lambda^{0,p+1}$. We further have ${\sf a} = {\cal H}{\sf a} + \Pi_{\delb} {\sf a} + \Pi_{{\delb}^\dagger}{\sf a} = {\sf a}_1 + \Pi_{\delb^\dagger}{\sf a}$ because of the condition $\delb^\dagger {\sf a} = 0$, and hence from (\ref{simplified_MC}) we find that
$\Pi_{\delb^\dagger}{\sf a} = -(1/\delb) \Pi_{\delb}[{\sf a},{\sf a}]$. Thus we can write
\be
{\sf a} + {1\over \delb}\Pi_{\delb}([{\sf a},{\sf a}]) = {\sf a}_1
\ee
We can solve this order by order. In fact if we define $T_{{\sf a}_1}(x) = {\sf a}_1 - {1\over \delb}\Pi_{\delb}([x,x])$ then ${\sf a}$ satisfies the fixed point equation $T_{{\sf a}_1}({\sf a}) = {\sf a}$. We can then define the sequence ${\sf s}_{n+1} = T_{{\sf a}_1}({\sf s}_n)$ where we defined the partial sums
${\sf s}_n = \sum_{k \leq n} {\sf a}_k$. The solution is known to converge in a small neighbourhood of the origin, i.e. if ${\sf a}_1$ is sufficiently small \cite{Fukaya:2001uc}.

Defining $g(y) = y +  {1\over \delb}\Pi_{\delb}([y,y])$, we see that if $a$ is constructed from $a_1$ as above then
$g(a) = a_1$, and this map is a local isomorphism. In other words, we have $g^{-1}(a_1) = a_1 + a_2 + \ldots$. Defining
the Kuranishi map $\kappa: H^{1}({\rm End}_0({\cal W})) \to H^{2}({\rm End}_0({\cal W}))$ as
\be
\label{Kuranishi_def}
\kappa({\sf a}_1) \ = \ {\sf H}^{0,2} ([g^{-1}({\sf a}_1), g^{-1}({\sf a}_1)])
\ee
then its non-vanishing is precisely the obstruction to promoting the infinitesimal deformation ${\sf a}_1$ to a finite deformation, i.e.
it is precisely the obstruction to solving $\delb {\sf a} + [{\sf a}, {\sf a}] = 0$ for the finite deformation.
 Locally the space of holomorphic bundles is described by $\kappa^{-1}(0)$ (and quotienting by any residual gauge symmetry).
By expanding this in our power series and collecting terms of the same order, we may write
\be
\kappa(a_1) \ = \ \sum_{k\geq 2} {\sf m}_k(a_1, \ldots ,a_1)
\ee
where ${\sf m}_k$ describes the terms of order $k$. The ${\sf m}_k$ satisfy the $A_\infty$ relations and the ${\sf m}_k$ are therefore the Massey products of an $A_\infty$ algebra.

Let us consider the leading terms explicitly. The leading non-zero term is
\be
{\sf m}_2(a_1, a_1) \ = \ {\cal H}^{0,2} ([{\sf a}_1,{\sf a}_1])
\ee
which is the usual Yoneda product $H^{0,1} \times H^{0,1} \to H^{0,2}$. Next we have
\be
{\sf a}_2 \ = \  - {1\over \delb} \Pi_{\delb}([{\sf a}_1,{\sf a}_1]
\ee
and
\be
{\sf m}_3 \ = \ {\cal H}^{0,2}([a_1, a_2] + [a_2, a_1]) \ = \ {\cal H}^{0,2}([a_1, {1\over \delb} \Pi_{\delb}([a_1,a_1]] + [{1\over \delb} \Pi_{\delb}([a_1,a_1], a_1])
\ee
This is equivalent to the classic Massey triple product.

Now we wish to consider the case of holomorphic ASD equations on a Calabi-Yau four-fold $Y$. As we discussed
in section \ref{Local_moduli}, by the results of \cite{Brav:2013} we want to pick a Kuranishi map
$\kappa$ for obstructions of holomorphic bundles such that $Q(\kappa,\kappa) = 0$. Fortunately
a short calculation in \cite{Cao:2014bca} shows that with $\kappa$ as defined above in (\ref{Kuranishi_def})
we indeed have
$Q(\kappa,\kappa) = 0$. Thus we can take the self-dual projection $\kappa_+ = \Pi_+ \circ\, \kappa$
to be a Kuranishi map for the complex ASD equations.\footnote{
We note that Kuranishi maps are not unique and Cao and Leung ultimately pick a different Kuranishi map denoted by $\tilde{\tilde{\kappa}}$. This allows them
to justify that the Kuranishi structures can be glued together globally over the moduli space. In our setting
we only need the local structure of the moduli space and hence $\kappa$ is fine for our purposes. We would like to thank
Y. Cao and N. Leung for correspondence on this issue.}


\newpage



\begin{thebibliography}{99}

\bibitem{Witten:1992fb}
  E.~Witten,
  ``Chern-Simons gauge theory as a string theory,''
  Prog.\ Math.\  {\bf 133}, 637 (1995)
  [hep-th/9207094].

\bibitem{Merkulov_homotopy}
S.A.~Merkulov,
``Strongly homotopy algebras of a K\"ahler manifold,''
 Internat. Math. Res. Notices (1999), no.3, 153--164,
 arXiv:math/9809172 [math.AG].

\bibitem{Polishchuk_HM}
A.~Polishchuk,
``Homological mirror symmetry with higher products,''
eprint arXiv:math/9901025

\bibitem{Kontsevich:2000yf}
  M.~Kontsevich and Y.~Soibelman,
  ``Homological mirror symmetry and torus fibrations,''
  math/0011041 [math-sg].

\bibitem{Lazaroiu:2001nm}
  C.~I.~Lazaroiu,
  ``String field theory and brane superpotentials,''
  JHEP {\bf 0110}, 018 (2001)
  doi:10.1088/1126-6708/2001/10/018
  [hep-th/0107162].

\bibitem{Kajiura:2005sn}
  H.~Kajiura and J.~Stasheff,
  ``Open-closed homotopy algebra in mathematical physics,''
  J.\ Math.\ Phys.\  {\bf 47}, 023506 (2006)
  doi:10.1063/1.2171524
  [hep-th/0510118].

\bibitem{Aspinwall:2009isa}
  P.~S.~Aspinwall {\it et al.},
  ``Dirichlet branes and mirror symmetry,''


\bibitem{Strominger:1995cz}
A.~Strominger,
``Massless black holes and conifolds in string theory,''
Nucl. Phys. B \textbf{451} (1995), 96-108
doi:10.1016/0550-3213(95)00287-3
[arXiv:hep-th/9504090 [hep-th]].

\bibitem{Vafa:1996xn}
  C.~Vafa,
  ``Evidence for F theory,''
  Nucl.\ Phys.\ B {\bf 469}, 403 (1996)
  [hep-th/9602022].

\bibitem{Morrison:1996na}
  D.~R.~Morrison and C.~Vafa,
  ``Compactifications of F theory on Calabi-Yau threefolds. 1,''
  Nucl.\ Phys.\ B {\bf 473}, 74 (1996)
  doi:10.1016/0550-3213(96)00242-8
  [hep-th/9602114].

\bibitem{Morrison:1996pp}
  D.~R.~Morrison and C.~Vafa,
  ``Compactifications of F theory on Calabi-Yau threefolds. 2.,''
  Nucl.\ Phys.\ B {\bf 476}, 437 (1996)
  doi:10.1016/0550-3213(96)00369-0
  [hep-th/9603161].

\bibitem{Becker:1996gj}
  K.~Becker and M.~Becker,
  ``M-Theory on Eight-Manifolds,''
  Nucl.\ Phys.\  B {\bf 477}, 155 (1996)
  [arXiv:hep-th/9605053].


\bibitem{Diaconescu:2003bm}
  E.~Diaconescu, G.~W.~Moore and D.~S.~Freed,
  ``The M theory three form and E(8) gauge theory,''
  hep-th/0312069.

\bibitem{Witten:1996md}
  E.~Witten,
  ``On flux quantization in M theory and the effective action,''
  J.\ Geom.\ Phys.\  {\bf 22}, 1 (1997)
  doi:10.1016/S0393-0440(96)00042-3
  [hep-th/9609122].


\bibitem{FM}
  R.~Friedman and J.~W.~Morgan,
  ``Exceptional groups and del Pezzo surfaces,''
  eprint arXiv:math/0009155.

\bibitem{ChenLeung}
  Y.~Chen and N.~C.~Leung,
  ``ADE bundles over surfaces with ADE singularities,''
	eprint arXiv:1209.4979[math.AG].

\bibitem{Baulieu:1997jx}
  L.~Baulieu, H.~Kanno and I.~M.~Singer,
  ``Special quantum field theories in eight-dimensions and other dimensions,''
  Commun.\ Math.\ Phys.\  {\bf 194}, 149 (1998)
  doi:10.1007/s002200050353
  [hep-th/9704167].

\bibitem{Baulieu:1997em}
  L.~Baulieu, H.~Kanno and I.~M.~Singer,
  ``Cohomological Yang-Mills theory in eight-dimensions,''
  In `Seoul/Sokcho 1997, Dualities in gauge and string theories' 365-373
  [hep-th/9705127].

\bibitem{Brav:2013}
  C.~Brav, V.~Bussi, and D.~Joyce,
  ``A 'Darboux theorem' for derived schemes with shifted symplectic structure,''
  arXiv:1305.6302

\bibitem{Cao:2014bca}
  Y.~Cao and N.~C.~Leung,
  ``Donaldson-Thomas theory for Calabi-Yau 4-folds,''
  arXiv:1407.7659 [math.AG].

\bibitem{Borisov:2015vha}
  D.~Borisov and D.~Joyce,
  ``Virtual fundamental classes for moduli spaces of sheaves on Calabi-Yau four-folds,''
  arXiv:1504.00690 [math.AG].

\bibitem{Freed:2019sco}
D.~S.~Freed and M.~J.~Hopkins,
``Consistency of M-Theory on Non-Orientable Manifolds,''
Quart. J. Math. Oxford Ser. \textbf{72}, no.1-2, 603-671 (2021)
doi:10.1093/qmath/haab007
[arXiv:1908.09916 [hep-th]].


 \bibitem{ADE_Transform}
 R.~Donagi and M.~Wijnholt,
 ``ADE Transform.''
Adv. Theor. Math. Phys. \textbf{24}, no.8, 2043-2066 (2020)
doi:10.4310/ATMP.2020.v24.n8.a2
[arXiv:1510.05025 [math.AG]].

\bibitem{Donagi:2011jy}
  R.~Donagi and M.~Wijnholt,
  ``Gluing Branes, I,''
  arXiv:1104.2610 [hep-th].

\bibitem{Marsano:2012bf}
  J.~Marsano, N.~Saulina and S.~Schafer-Nameki,
  ``Global Gluing and $G$-flux,''
  JHEP {\bf 1308}, 001 (2013)
  doi:10.1007/JHEP08(2013)001
  [arXiv:1211.1097 [hep-th]].

\bibitem{Collinucci:2014taa}
  A.~Collinucci and R.~Savelli,
  ``F-theory on singular spaces,''
  arXiv:1410.4867 [hep-th].

\bibitem{Anderson:2013rka}
  L.~B.~Anderson, J.~J.~Heckman and S.~Katz,
  ``T-Branes and Geometry,''
  JHEP {\bf 1405}, 080 (2014)
  doi:10.1007/JHEP05(2014)080
  [arXiv:1310.1931 [hep-th]].

\bibitem{MF_Extension}
 R.~Donagi and M.~Wijnholt, unpublished.

\bibitem{F_Fluxes}
 R.~Donagi and M.~Wijnholt, unpublished.

\bibitem{F_Stability}
 R.~Donagi and M.~Wijnholt,
 ``Stability conditions for the $M$-theory three-form,'' unpublished.


\bibitem{Brooks:1988jm}
R.~Brooks, D.~Montano and J.~Sonnenschein,
``Gauge Fixing and Renormalization in Topological Quantum Field Theory,''
Phys. Lett. B \textbf{214}, 91-97 (1988)
doi:10.1016/0370-2693(88)90458-3

\bibitem{Baulieu:1988xs}
  L.~Baulieu and I.~M.~Singer,
  ``Topological Yang-mills Symmetry,''
  Nucl.\ Phys.\ Proc.\ Suppl.\  {\bf 5B}, 12 (1988).
  doi:10.1016/0920-5632(88)90366-0

\bibitem{LeungADE}
  Leung, Naichung Conan,
  ``ADE-bundles over rational surfaces, configuration of lines and rulings,''
  arXiv:math/0009192.

\bibitem{LZ1}
  Leung, Naichung Conan; Zhang, Jiajin,
  ``Moduli of bundles over rational surfaces and elliptic curves I: simply laced cases,''
   J. Lond. Math. Soc. (2) 80 (2009), no. 3, 750-770,
  arXiv:0906.3900.

\bibitem{LZ2}
    Leung, Naichung Conan; Zhang, Jiajin,
    ``Moduli of bundles over rational surfaces and elliptic curves II: non-simply laced cases,''
    Int. Math. Res. Not. IMRN 2009, no. 24, 4597-4625,
    arXiv:0908.1645.

\bibitem{LeungChenFlag}
Y.~Chen, N.~C.~Leung,
``ADE Bundles Over ADE Singular Surfaces and Flag Varieties of ADE Type,''
IMRN: International Mathematics, 2018.
arXiv preprint arXiv:1811.02777.

\bibitem{Friedman:1997ih}
  R.~Friedman, J.~W.~Morgan and E.~Witten,
  ``Vector bundles over elliptic fibrations,''
  alg-geom/9709029.

\bibitem{Horava:1996ma}
P.~Horava and E.~Witten,
``Eleven-dimensional supergravity on a manifold with boundary,''
Nucl. Phys. B \textbf{475}, 94-114 (1996)
doi:10.1016/0550-3213(96)00308-2
[arXiv:hep-th/9603142 [hep-th]].

\bibitem{Witten:2019bou}
E.~Witten and K.~Yonekura,
``Anomaly Inflow and the $\eta$-Invariant,''
[arXiv:1909.08775 [hep-th]].


\bibitem{FriedmanMorgan}
  R.~Friedman and J.~Morgan,
  ``Minuscule representations, invariant polynomials, and spectral covers,''	
  eprint arXiv:math/0011082[math.AG].

\bibitem{Donagi:2010pd}
  R.~Donagi and M.~Wijnholt,
  ``MSW Instantons,''
  JHEP {\bf 1306}, 050 (2013),
  [arXiv:1005.5391 [hep-th]].

\bibitem{Wijnholt:2012fx}
  M.~Wijnholt,
  ``Higgs Bundles and String Phenomenology,''
  Proc. Symp. Pure Math. \textbf{85}, 275-292 (2012)
  doi:10.1090/pspum/085/1383
  [arXiv:1201.2520 [math.AG]].


\bibitem{DonagiCovers}
 R.~Donagi,
 ``Principal bundles on elliptic fibrations,''
 Asian J. Math. Vol. 1 (June 1997), 214-223.
 [arXiv:alg-geom/9702002].

\bibitem{Curio:1998bva}
G.~Curio and R.~Y.~Donagi,
``Moduli in N=1 heterotic / F theory duality,''
Nucl. Phys. B \textbf{518}, 603-631 (1998)
doi:10.1016/S0550-3213(98)00185-0
[arXiv:hep-th/9801057 [hep-th]].

\bibitem{Aker_dp_fibrations}
K.~Aker,
``Almost Regular Bundles on del Pezzo Fibrations,''
arXiv preprint math/0508557.

\bibitem{UniTorsors_2016}
U.~Derenthal, N.~Hoffmann,
``Degeneration of torsors over families of del Pezzo surfaces,''
Mathematische Nachrichten, 2020,
arXiv preprint arXiv:1601.00685.

\bibitem{Clingher:2012rg}
  A.~Clingher, R.~Donagi and M.~Wijnholt,
  ``The Sen Limit,''
  Adv.\ Theor.\ Math.\ Phys.\  {\bf 18}, 613 (2014)
  [arXiv:1212.4505 [hep-th]].


\bibitem{Hayashi:2008ba}
H.~Hayashi, R.~Tatar, Y.~Toda, T.~Watari and M.~Yamazaki,
``New Aspects of Heterotic--F Theory Duality,''
Nucl. Phys. B \textbf{806}, 224-299 (2009)
doi:10.1016/j.nuclphysb.2008.07.031
[arXiv:0805.1057 [hep-th]].

\bibitem{Donagi:2008ca}
  R.~Donagi and M.~Wijnholt,
  ``Model Building with F-Theory,''
  arXiv:0802.2969 [hep-th].

\bibitem{Kachru:1999vj}
S.~Kachru and J.~McGreevy,
``Supersymmetric three cycles and supersymmetry breaking,''
Phys. Rev. D \textbf{61}, 026001 (2000)
doi:10.1103/PhysRevD.61.026001
[arXiv:hep-th/9908135 [hep-th]].

\bibitem{Haack:2002tu}
M.~Haack,
``Calabi-Yau fourfold compactifications in string theory,''
Fortsch. Phys. \textbf{50}, 3-99 (2002)

\bibitem{Donagi:2008kj}
  R.~Donagi and M.~Wijnholt,
  ``Breaking GUT Groups in F-Theory,''
  Adv.\ Theor.\ Math.\ Phys.\  {\bf 15}, 1523 (2011)
  [arXiv:0808.2223 [hep-th]].


\bibitem{Witten:1997bs}
E.~Witten,
``Toroidal compactification without vector structure,''
JHEP \textbf{02}, 006 (1998)
doi:10.1088/1126-6708/1998/02/006
[arXiv:hep-th/9712028 [hep-th]].


\bibitem{deBoer:2001wca}
J.~de Boer, R.~Dijkgraaf, K.~Hori, A.~Keurentjes, J.~Morgan, D.~R.~Morrison and S.~Sethi,
``Triples, fluxes, and strings,''
Adv. Theor. Math. Phys. \textbf{4}, 995-1186 (2002)
doi:10.4310/ATMP.2000.v4.n5.a1
[arXiv:hep-th/0103170 [hep-th]].


\bibitem{Gukov:1999ya}
  S.~Gukov, C.~Vafa and E.~Witten,
  ``CFT's from Calabi-Yau four folds,''
  Nucl.\ Phys.\ B {\bf 584}, 69 (2000)
  Erratum: [Nucl.\ Phys.\ B {\bf 608}, 477 (2001)]
  doi:10.1016/S0550-3213(01)00289-9, 10.1016/S0550-3213(00)00373-4
  [hep-th/9906070].

\bibitem{Witten:1988ze}
  E.~Witten,
  ``Topological Quantum Field Theory,''
  Commun.\ Math.\ Phys.\  {\bf 117}, 353 (1988).
  doi:10.1007/BF01223371

\bibitem{Acharya:1997jn}
  B.~S.~Acharya, J.~M.~Figueroa-O'Farrill, B.~J.~Spence and M.~O'Loughlin,
  ``Euclidean D-branes and higher dimensional gauge theory,''
  Nucl.\ Phys.\ B {\bf 514}, 583 (1998)
  doi:10.1016/S0550-3213(97)00727-X
  [hep-th/9707118].

\bibitem{Bak:2002aq}
  D.~s.~Bak, K.~M.~Lee and J.~H.~Park,
  ``BPS equations in six-dimensions and eight-dimensions,''
  Phys.\ Rev.\ D {\bf 66}, 025021 (2002)
  doi:10.1103/PhysRevD.66.025021
  [hep-th/0204221].

\bibitem{Donaldson:1996kp}
  S.~K.~Donaldson and R.~P.~Thomas,
  ``Gauge theory in higher dimensions,''
  The Geometric Universe: Science, Geometry and the work of Roger Penrose, (SA
 Huggett et al eds.), Oxford University Press, 1998.

\bibitem{ThomasGaugeTheory}
  R.~P.~Thomas,
  ``Gauge Theory on Calabi-Yau Manifolds,
  Ph.D. Thesis, 1997. \newline
   R.~P.~Thomas,
  ``A holomorphic Casson invariant for Calabi-Yau 3-folds, and bundles on K3
  fibrations,''
  eprint arXiv:math/9806111,
  Jour. Diff. Geom. 54, no. 2, 367-438, 2000.

 \bibitem{Pantev:2009de}
  T.~Pantev and M.~Wijnholt,
  ``Hitchin's Equations and M-Theory Phenomenology,''
  J.\ Geom.\ Phys.\  {\bf 61}, 1223 (2011)
  [arXiv:0905.1968 [hep-th]].

\bibitem{Lewis:1998}
  C.~Lewis,
  ``Spin(7) Instantons,''
  DPhil. Oxford University, 1998.

\bibitem{Stasheff:1997iz}
J.~Stasheff,
``The (Secret?) homological algebra of the Batalin-Vilkovisky approach,''
Contemp. Math. \textbf{219}, 195-210 (1998)
doi:10.1090/conm/219/03076
[arXiv:hep-th/9712157 [hep-th]].

\bibitem{Costello:2016vjw}
K.~Costello and O.~Gwilliam,
``Factorization Algebras in Quantum Field Theory, vol. 2,''
Cambridge University Press 2016.

\bibitem{Oh:2020rnj}
J.~Oh and R.~P.~Thomas,
``Counting sheaves on Calabi-Yau 4-folds, I,''
[arXiv:2009.05542 [math.AG]].

\bibitem{Witten:1993yc}
E.~Witten,
``Phases of N=2 theories in two-dimensions,''
Nucl. Phys. B \textbf{403}, 159-222 (1993)
doi:10.1016/0550-3213(93)90033-L
[arXiv:hep-th/9301042 [hep-th]].

\bibitem{Erler:2013xta}
  T.~Erler, S.~Konopka and I.~Sachs,
  ``Resolving Witten`s superstring field theory,''
  JHEP {\bf 1404}, 150 (2014)
  doi:10.1007/JHEP04(2014)150
  [arXiv:1312.2948 [hep-th]].

\bibitem{Baulieu:2010ch}
L.~Baulieu,
``SU(5)-invariant decomposition of ten-dimensional Yang-Mills supersymmetry,''
Phys. Lett. B \textbf{698} (2011), 63-67
doi:10.1016/j.physletb.2010.12.044
[arXiv:1009.3893 [hep-th]].

\bibitem{Costello:2016mgj}
K.~Costello and S.~Li,
``Twisted supergravity and its quantization,''
[arXiv:1606.00365 [hep-th]].

\bibitem{TianYau_I}
 G.~Tian and S.~T.~Yau,
 ``Complete K\"ahler manifolds with zero Ricci curvature, I,''
  J. Amer. Math. Soc. {\bf 3} (1990), no. 3, 579-609, MR 1040196 (91a:53096).

\bibitem{TianYau_II}
 G.~Tian and S.~T.~Yau,
 ``Complete K\"ahler manifolds with zero Ricci curvature, II,''
  Invent. Math {\bf 106} (1991), no. 1, 27-60, MR 1123371 (92j:32028).

\bibitem{Kovalev:2000}
 A.~Kovalev,
 ``Twisted connected sums and special Riemannian holonomy,''
  J. Reine Angew. Math.565(2003),125-160,
  arXiv:math/0012189 [math.DG].

\bibitem{ACyl_CY:2012}
 M.~Haskins, H.-J.~Hein and J.~Nordstr\"om,
 ``Asymptotically cylindrical Calabi-Yau manifolds,
 J. Differential Geom. Volume 101, Number 2 (2015), 213-265,
 arXiv: 1212.6929[math.DG].

\bibitem{Melrose_APS}
 R.~B.~Melrose,
 ``The Atiyay-Patodi-Singer index theorem,''
 A.K.Peters, Wellesley, Massachusetts, 1993.

\bibitem{ACyl_moduli:2014}
 R.~J.~Conlon, R.~Mazzeo and F.~Rochon,
 ``The moduli space of asymptotically cylindrical Calabi-Yau manifolds,''
  Commun. Math. Phys. (2015) 338: 953,
 arXiv: 1408.6562[math.DG].

\bibitem{Donaldson:2002}
 S.~K.~Donaldson,
 ``Floer homology groups in Yang-Mills theory,''
 Cambridge University Press, 2002.

\bibitem{Collins:2019}
 T.~C.~Collins, A.~Jacob and Y.-S.~Lin,
 ``Special Lagrangian submanifolds of log Calabi-Yau manifolds,''
 arXiv: 1904.08363 [math.DG]

\bibitem{Huang:1995}
Huang, Liaw. "On joint moduli spaces.." Mathematische Annalen 302.1 (1995): 61-80. 

\bibitem{ChanSuen:2014}
 Chan, K. and Suen, Y.-H., 
 “A differential-geometric approach to deformations of pairs $(X,E)$”, 
 doi:10.48550,
 arXiv.1406.6753.
 
\bibitem{Iacono:2013}
Iacono, D.: 2013, 
 “Deformations and obstructions of pairs (X,D),”
 {\it arXiv e-prints}, arXiv:1302.1149. doi:10.48550/arXiv.1302.1149.


\bibitem{Katzarkov:2014bma}
L.~Katzarkov, M.~Kontsevich and T.~Pantev,
``Bogomolov-Tian-Todorov theorems for Landau-Ginzburg models,''
J. Diff. Geom. \textbf{105}, no.1, 55-117 (2017)
[arXiv:1409.5996 [math.AG]].


\bibitem{Fukaya:2001uc}
  K.~Fukaya,
  ``Deformation theory, homological algebra and mirror symmetry,''

\end{thebibliography}
\end{document}